\documentclass[12pt]{article}
\usepackage{url}
\usepackage{t1enc}
\usepackage{amsmath}
\usepackage{amsthm}
\usepackage{latexsym}
\usepackage{epsfig}
\usepackage{textcomp}
\usepackage{alltt}
\usepackage{graphicx}
\usepackage{rotating}
\usepackage{dcolumn}
\usepackage{enumitem}
\usepackage{amsfonts}
\usepackage{algorithm}
\usepackage{amssymb}
\usepackage{bbm, dsfont}
\usepackage[svgnames]{xcolor}
\usepackage{mathtools}
\usepackage{listings}
\usepackage{pdflscape}
\usepackage{afterpage}
\usepackage{capt-of}

\usepackage{tikz}
\tikzstyle{ov}=[shape=rectangle,
                draw=black!50,
                thick,
                minimum width=0.7cm,
                minimum height=0.7cm]

\tikzstyle{av}=[shape=rectangle,
                draw=black!50,
                fill=black!10,
                thick,
                minimum width=0.7cm,
                minimum height=0.7cm]

\tikzstyle{lv}=[shape=circle,draw=black!50,thick]

\allowdisplaybreaks

\begin{document}

\parindent12pt

\begin{center}{\Large{Estimating the optimal time to perform a PET-PSMA
  exam in prostatectomized patients based on data from clinical practice}}
 \end{center}
 
 { 
\begin{center}
Martina Amongero$^{1}$, Gianluca Mastrantonio$^{2}$, Stefano De Luca$^{3}$, Mauro Gasparini$^{2}$\\
\bigskip
\scriptsize{$^1$Department of Economics, Social Studies, Applied Mathematics, and Statistics, {Università di Torino}, Torino, Italy} \\
\scriptsize{$^2$Department of Mathematical Sciences, Politecnico di Torino, Torino, Italy} \\
\scriptsize{$^3$Department of Urology, San Luigi Hospital, Torino, Italy} 
\end{center}
}
\smallskip
\begin{center}
\today
\end{center}
\smallskip

\setlength{\parindent}{0.3in} \setlength{\baselineskip}{24pt}

\begin{abstract}
Prostatectomized patients are at risk of resurgence, and for this reason, during a follow-up period, they are monitored for Prostate Specific Antigen (PSA) growth, an indicator of tumor progression. The presence of tumors can be evaluated with an expensive exam, called Positron Emission Tomography with Prostate-Specific Membrane Antigen (PET-PSMA).
To justify the high cost of the PET-PSMA and, at the same time, to contain the risk for the patient, this exam should be recommended only when the evidence of tumor progression is strong.  
With the aim of estimating the optimal time to recommend the exam based on the patient's history and collected data, 
we build a hierarchical Bayesian model that describes, jointly, the PSA growth curve and the probability of a positive PET-PSMA. With our proposal we process all past and present information about the patients PSA measurement and PET-PSMA results, in order to give an informed estimate of the optimal time, improving current practice.
\end{abstract}







\section{Introduction}\label{literaturePSA}
Nowadays, one of the most important areas of medical statistical development is oncology. The statistical and clinical interest varies from the early detection of tumor presence and locations, the estimation of (personalized) treatment efficacy, the definition of optimal personalized treatment strategies, the analyses of resurgences, the development of ethical trial structures, to many other problems.


Prostate cancer is the most frequent neoplasm in men, with an incidence of around 7\% among all new cancer cases \cite{cancer}, and has therefore attracted a lot of interest in the last twenty years.
Research focuses on several  topics, starting from the causes and the incidence of the tumor, modeling its growth,  analyzing the effect of therapies, and, finally, analyzing the risk, the locations, and time of biochemical relapse (BCR) and clinical recurrences.
High levels of serum Prostate-Specific Antigen (PSA) after primary treatments, such as surgery or radiotherapy, were identified in the literature to be significant indicators of tumor progression taking place at some locations different from the original one
\cite{148,150,151,149,152}.
This means that prostatectomized patients are at risk of developing BCR and metastasis, which is the reason why, during a follow-up period of several years, they are usually monitored for PCA recurrence, traditionally identified by PSA $>0.2$ ng/mL.

 Positron Emission Tomography with Prostate-Specific Membrane Antigen (PET-PSMA) is a nuclear medicine survey, that is currently one of the most sensitive tests for the early detection of tumor presence and location.
It is very expensive and complex and, for this reason, patients should be referred to a PET-PSMA exam only in case of confirmed high risk of BCR.
The timing of the exam is very important since if it is performed too late, the patient is subject to excessive risks, while, if it is too soon,  there is a high probability of false negative results.
The estimation of the optimal time to perform such an examination is still an open problem widely discussed in the literature.

Currently, the most used indicator of BCR is the  PSA Doubling Time (PSA-DT) which is usually monitored over time to control the PSA evolution \cite{PSADT6month}.
This indicator is calculated using only the last two measurements available for the patient analyzed.
According to literature \cite{PSADT6month}, a PSA-DT$<$6 months, together with a thorough clinical examination of the patient stage and conditions, should be used as an indication to perform a PET-PSMA exam.
Some authors model the PSA evolution over time trying to link it to tumor progression \cite{slate1997changepoint,hirata2012quantitative}, distinguishing between patients with and without disease \cite{carter1992longitudinal,proust2009development}, or accounting of the impact of aging on its evolution \cite{pearson1991modeling}.
Other work mainly focuses on the PET examination \cite{fossati2019emerging} or on the correlation between PSA levels and PET results \cite{pereira2019correlation}.
Finally, some researchers link PSA kinetics and tumor recurrence \cite{proust2009development}, whereas other authors try to determine the optimal time of PET-PSMA \cite{luiting2020optimal} regardless of the kinetics of PSA.
Our work combines many aspects of all these papers into an overall joint statistical model \cite{verbeke,joint,duchateau2008frailty}, and exploits its structure to make predictions. While other approach propose to model either  the PSA  and the potential resurgences (see, for example, \cite{slate1997changepoint,hirata2012quantitative,pearson1991modeling}), or the predicting the probability of PET-PSMA positive results (\cite{fossati2019emerging,luiting2020optimal}), we propose a joint model that combines both aspects to improve the accuracy of the estimates.
\textcolor{black}{We base our model construction on the joint-model idea proposed by \cite{proust2009development}, in which the authors model the risk of resurgence for a cohort of patients who underwent radiotherapy. They assume that patients after radiotherapy can be divided into classes and that PSA and the risk of recurrence are independent given the latent class membership. The probability that a subject belongs to a class is related to the covariates through a multinomial logistic regression model, with parameters estimated using the maximum likelihood approach. 
Differently from \cite{proust2009development}, we link the PSA and the results of the PET-PSMA through a latent description of the true PSA level, which is defined as a patient-specific functional form. Our aims are also different since, by analyzing a dataset from an observational study on prostatectomized patients, we want to develop a methodology that can be generalized to improve the standard of care and be used to predict and provide a measure of uncertainty regarding the optimal time to perform the PET-PSMA exam. }

\textcolor{black}{We assume that the true, yet unobservable, PSA levels depend on patient covariates 
and on the presence of BCR. Specifically, as in \cite{proust2009development}, the latent PSA level is expected to decrease before BCR and to increase afterward.
Both the measured PSA level and PET-PSMA results are dependent on this latent structure, which is probabilistically linked with them.
One of the significant advantages of our approach is that we are utilizing all available information jointly to determine the presence of BCR and the likelihood of a positive PET-PSMA.
Moreover, in contrast to other methods that link PET-PSMA results to the measured PSA levels, our approach allows for estimating the
probability of a positive PET-PSMA without an associated PSA measurement.
By selecting a target probability of having a positive PET-PSMA and a reliability level, we can estimate the optimal time to perform the test with Markov Chain Monte Carlo methods.  Inference is performed according to the principles of Bayesian statistics, which allows us to have a complete evaluation of the uncertainty associated with each model component.
To do so, a Bayesian network \cite{BayesianNetwork} is built,
or more precisely a Bayesian hierarchical model.
Bayesian networks are probabilistic graphical models which represent
a set of variables and their probabilistic interdependencies,
allowing for the data analyst to have an accurate uncertainty quantification
of all model unknown parameters.  The dataset used in this paper is part of a  collaboration with the clinicians of the Department of Urology of San Luigi Gonzaga Hospital in Torino (Italy).}

The paper is divided as follows: in Section \ref{datasection} we present the joint model for PSA measurements and PET-PSMA results.
Section \ref{sec:optimla} is devoted to the definition and the estimation of the optimal time to perform a test.
We test our model on simulated data, as explained in Section \ref{sec:sim}, before applying it to real data: Section \ref{sec:real} contains the results of the model estimated on a group of patients from San  Luigi Gonzaga Hospital, in Torino, Italy. The paper ends with some conclusive remarks in Section \ref{sec5}.

\section{Joint model of PSA growth curve and PET-PSMA results}\label{datasection}

This Section is devoted to explain the statistical joint model: we first describe the growth model for the PSA curve (Sub-section \ref{sec:ss1}), and then how this model is linked to the logistic structure for the result of PET-PSMA (Sub-section \ref{jointmodelsection}).
Finally, Sub-section \ref{sec:ss3} deals with random effects.

\subsection{The data and its likelihood} \label{sec:ss1}

Let $x_i(t)$ be the PSA level of the $i$-th patient 
at time $t$, where $t=0$  indicates the time the patient has had a prostatectomy (only operated patients are considered in this work) and $i=1,2, \ldots, n$, where $n$ is the total number of patients in the study.
The recorded variable $y_i(t)$ is a noisy version of the non-negative quantity $x_i(t)$, therefore we assume
\begin{equation}
    \log y_i(t) \sim \mathcal{N}( \log x_i(t), \sigma_i^2).
\end{equation}
The choice of the functional form of $x_i(t)$, i.e., how PSA levels depend on time and other covariates, is one of the two major components of our proposed joint model (described in Sub-section \ref{jointmodelsection}), but before we discuss it, let us introduce the second component, which is the probability $\pi_i(t)$ for patient $i$ at time $t$ that, if a PET-PSMA is taken, the result will be positive. For each patient, the test outcome  $z_i(t) \in \{0,1\}$ is assumed to be Bernoulli random variable with a patient-specific and time-dependent probability, i.e.,
\begin{equation}
    z_i(t) \sim \textit{Ber}( \pi_i(t)).
\end{equation}
Variables $x_i(t)$ and $\pi_i(t)$ are not observed,  and we can only observe  $y_i(t)$ and $z_i(t)$ at specific time points. The time points where these two variables are recorded can differ within and between patients, and we indicate as  $\mathcal{T}_{y,i}$ and $\mathcal{T}_{z,i}$ the set of time points of patient $i$ where $y_i(t)$ and $z_i(t)$ are recorded, respectively.
The set  ${\mathcal{T}_i = \mathcal{T}_{y,i} \cup \mathcal{T}_{z,i}}$ is the set of points where we have a measure of at least one of the two variables.

Let
$\mathbf{y}_i^{o} = \{y_i(t)\}_{t \in \mathcal{T}_{y,i}}$,
$\mathbf{x}_i^{o} = \{x_i(t)\}_{t \in \mathcal{T}_{y,i}}$ be the observed and latent PSA values over the time points in $\mathcal{T}_{y,i}$ and let
$\mathbf{z}_i^{o} = \{z_i(t)\}_{t \in \mathcal{T}_{z,i}}$,
${\boldsymbol{\pi}_i^{o} = \{ {\pi}_i(t)\}_{t \in \mathcal{T}_{z,i}}}$ be the test results and probabilities over $\mathcal{T}_{z,i}$  ({i.e., multiple tests for the same patient can be collected, at multiple times, that do not necessarily correspond to times PSA measurements are collected}).
Conversely, using the superscript $u$ for ``unobserved'' (as opposed to the previous $o$ for ``observe''), let
$\mathbf{y}_i^{u} = \{y_i(t)\}_{t \in \mathcal{T}_{z,i}}$,
$\mathbf{x}_i^{u} = \{x_i(t)\}_{t \in \mathcal{T}_{z,i}}$,
$\mathbf{z}_i^{u} = \{z_i(t)\}_{t \in \mathcal{T}_{y,i}}$,
$\boldsymbol{\pi}_i^{u} = \{ {\pi}_i(t)\}_{t \in \mathcal{T}_{y,i}}$ the vectors of variables at the time points where the associated process is not measured, i.e., $(y_i(t), x_i(t))$ at time points  $\mathcal{T}_{z,i}$ where the exams are taken, and $(z_i(t), \pi_i(t))$ at time points  $\mathcal{T}_{y,i}$ where the PSA is measured, respectively.

High values of PSA are indicators of tumor progression and are thus associated with a high probability of positive PET-PSMA results. We will then assume that $\pi_i(t)$ is a function of $x_i(t)$, and use this relation to find the optimal PET-PSMA time since the central idea of this work is to exploit the joint model structure \cite{joint,proust2009development} to describe the available longitudinal data. To specify the model we have to define the joint density of variables $(\log y_i(t),z_i(t))'$  over $\mathcal{T}_i$, for both observed and missing data. Hence, the missing data are considered further parameters to be estimated during the model fitting, which are easy to estimate within the Bayesian framework.

The joint density for a random sample is then factorized in the following way
 \begin{equation} \label{eq:ind}
 \begin{split}
f\big(\log \mathbf{y}^u,\log \mathbf{y}^o, \mathbf{z}^u,\mathbf{z}^o\,&|\, \log \mathbf{x}^u,\log \mathbf{x}^o, \boldsymbol{\pi}^u,\boldsymbol{\pi}^o  ,\boldsymbol{\theta}\big) =\\ 
&\prod_{i=1}^n
\prod_{t \in \mathcal{T}_{i}} f\big(\log y_{i}(t)\,|\,\log x_{i}(t), \boldsymbol{\theta}\big) f\big(z_{i}(t)\,|\, \log x_{i} (t), \pi_{i}(t),\boldsymbol{\theta}\big),
\end{split}
\end{equation} 
where $f(\cdot)$ stands for a generic probability density function (to be identified by its arguments) and $\boldsymbol{\theta}$ is a vector of parameters. In Equation \eqref{eq:ind} we are assuming that $(\log y_{i} (t), z_{i}(t))$ are conditionally independent given the latent variables since the connection between the two measurements $y_i(t)$  and $z_i(t)$ is through the latent PSA level $x_i(t)$, and once we know $x_i(t)$, no further information is needed to explain the PET-PSMA result.  In the next paragraphs we illustrate our proposals for $f(\log y_{i} (t)\,|\,\log x_{i} (t), \boldsymbol{\theta})$ and  $f(z_{i}(t)\,|\, \log x_{i} (t), \pi_{i}(t),\boldsymbol{\theta})$. We want to remark that in Equation \eqref{eq:ind} we are assuming conditional independence between the measurements, but, on the other hand,  as we show in Sub-section \ref{sec:ss3}, we introduce random effects over some components of $\boldsymbol{\theta}$, to model  {dependencies across patients}.

\subsection{Joint model for each patient} \label{jointmodelsection}
To model the time evolution of PSA levels, we assume that it is composed of two phases, the first one, right after prostatectomy, where the PSA level is stationary or even decreasing over time, and the second one, after a patient-dependent change point in time $\tau_i$, where it is assumed that the PSA increases, until reaching a plateau  indicated as $a_i$. 
For each patient, the time $\tau_i$ can be interpreted as the unknown time at which resurgence starts, which is a crucial object of inference in our model.
The component $\log x_i(t)$ is modeled as a linear function with patient-specific intercept
$\lambda_i$ and slope $-\mu_i$, for non-negative $\mu_i$, i.e., if $t\leq\tau_i$ 
\begin{equation}\label{eq:logx1}
\log x_i(t) = \lambda_i- \mu_{i}t,
\end{equation}
while for $t>\tau_i$, we model  $\log x_i (t_i)$ \textcolor{black}{using a log-Gompertz growth model, which is a standard functional form to describe cancer and cancer-related biomarker evolution  \cite{gompertz,gompertsfortumor}. By the log-Gompertz model, the evolution function is a weighted mean of the value assumed by $\log x_i (t)$ at the change point, i.e., $\log x_i (\tau_i)$, and its asymptotic value $a_i$:}
\begin{equation} \label{eq:logx}
\log x_i(t) = \log x_i (\tau_i) e_i(t)+a_i (1-e_i(t)).
\end{equation}
The weight function  $e_i(t)$ is defined as
\begin{equation}
e_i(t)=\exp(-\gamma_i(t-\tau_i)).
\end{equation}
 Hence, for $t> \tau_i$,  $x_i(t)$ is a version of the
log-Gompertz growth function \cite{gompertz,gompertsfortumor} with rate $\gamma_i\in \mathbb{R}^+$ and asymptote $a_i\in \mathbb{R}^+$.
We assume non-negative $\mu_i$ and $\gamma_i$ to model the decreasing phase before $\tau_i$ and the increasing phase after it.

The second part of the joint model is instead focused on the binary PET-PSMA results $ z_{i}(t)$,
and in particular on modeling its probability $\pi_i(t)$. To connect the probability $\pi_i(t)$ to the PSA levels we use a logistic regression model
\begin{equation}\label{eq:logit}
     \text{logit}\{\pi_i(t)\} = \beta_{0,i} + \beta_1 \log x_i(t)+  \beta_2 t,
\end{equation}
where the linear prediction has a patient-specific intercept
and an extra term linear on time to model temporal trends that are not explained by $\log x_i(t)$.
It should be noted that we define the relation in terms of the true latent PSA levels $x_i(t)$, not the observed ones $y_i(t)$.
We expect $\beta_1$ to be positive since the larger the PSA, 
the larger the probability of a positive test. It is easy to see, from Equation \eqref{eq:logit}, that, if $\beta_1>0$, then $\pi_i(t)$ goes to zero as $x_i(t)$ goes to zero:
\begin{equation}\label{eq:lim}
\lim_ {x_i(t) \to 0^+} \pi_i(t)= \lim_ {x_i(t) \to 0^+}  \frac{x_i^{\beta_1}(t) e^{ \beta_{0,i}+\beta_2 t}}{1+x_i^{\beta_1}(t) e^{ \beta_{0,i}+\beta_2 t}}=0.
\end{equation}
All the unknown quantities in the model $\{(\lambda_i, \mu_i, a_i, \gamma_i, \tau_i, \beta_{0,i}, \beta_1, \beta_2, \sigma_i^2)\}_{i = 1, \dots, n}$ make up the parameter vector $\boldsymbol{\theta}$.

\subsection{Random effects} \label{sec:ss3}

We now extend the base model, where each patient has its own set of parameters, to a random effects model.  Generally, random effects can help to account for variability and heterogeneity in the data, due to unobserved or unmeasured factors, and may lead to more accurate and reliable estimates of treatment effects.

More precisely,  we assume that the model describing PSA evolution can be enriched by the following second-level distributions:
\begin{equation}
\label{eq:re}
\begin{split} 
\log \mu_i & \sim \mathcal{N}(\psi_{i,\mu}, w_{\mu}^2),\\
\log \gamma_i & \sim \mathcal{N}(\psi_{i,\gamma}, w_{\gamma}^2),\\
a_i & \sim \mathcal{N}(\psi_{i,a}, w_{a}^2),\\
\lambda_i & \sim \mathcal{N}(\psi_{i,\lambda}, w_{\lambda}^2),\\
\sigma_i^2 & \sim \mathcal{IG}(a_{i,\sigma^2}, b_{i,\sigma^2}),
\end{split}
\end{equation}
where $\mathcal{IG}$ indicates the inverse gamma distribution with scale parameter $a_{i,\sigma^2}$ and shape parameter $b_{i,\sigma^2}$, and $\psi_{i,\chi}$ and $w_{\chi}^2$ are respectively mean and variance of the normal distributions, for $\chi \in \{\mu, \gamma,  a, \lambda\}$. The log transformations are used to ensure that the parameters are defined in the correct domain.
In Equation~\eqref{eq:re}, the means of the normal distributions are allowed to vary from patient to patient since there may exist covariates that affect them via a linear regression  as follows,
\begin{equation}
\psi_{i,\chi} = \mathbf{C}_{i, \chi}\boldsymbol{\alpha}_{\chi},
\end{equation}
where $\mathbf{C}_{i, \chi}$ is a patient-specific vector of covariates of dimension $p_{\chi}$ and $\boldsymbol{\alpha}_{\chi}$ is a vector of regressors. 
Finally, covariate information can also be added to the component of the model that is used to define $\pi_i(t)$ by adding, in Equation \eqref{eq:logit}, the following random intercept,
 \begin{equation} \label{eq:b01}
\beta_{0,i} = \mathbf{C}_{i,\beta}\boldsymbol{\alpha}_{\beta}.
\end{equation}
 The random effects $(a_{i,\sigma^2}, b_{i,\sigma^2},\psi_{i,\chi}, w_{i,\chi}^2)$, for $\chi \in \{\mu, \gamma,  a, \lambda \}$, as well as the hyper-parameters $\boldsymbol{\alpha}_{\chi}$,  $\chi \in \{\mu, \gamma,  a, \lambda, \beta \}$  are all appended to the parameter vector $\boldsymbol{\theta}$.



\section{Estimating the optimal time} \label{sec:optimla}
Equation \eqref{eq:logit} links the probability of a positive PET-PSMA test and the latent PSA level; this relation between the two is what we use to find the optimal time. 
We recall that the PSA level (when observed) is not the true underlying level, but a noisy version of the true latent one that cannot be directly measured.

\subsection{Defining and identifying the optimal time}

{A solution to find the optimal time could be to plug in point estimates of all model parameters (maximum likelihood estimates, or Bayesian posterior means), and then invert Equation \eqref{eq:logit} to target the desired probability}.
This strategy is viable, but it does not take into account all sources of uncertainty.  The Bayesian approach, which we follow globally to fit the model, gives a better way to estimate the optimal time and, simultaneously, account for uncertainty quantification in a controlled way.

\begin{figure}[H]
  \centering
 \includegraphics[width=0.8\textwidth]{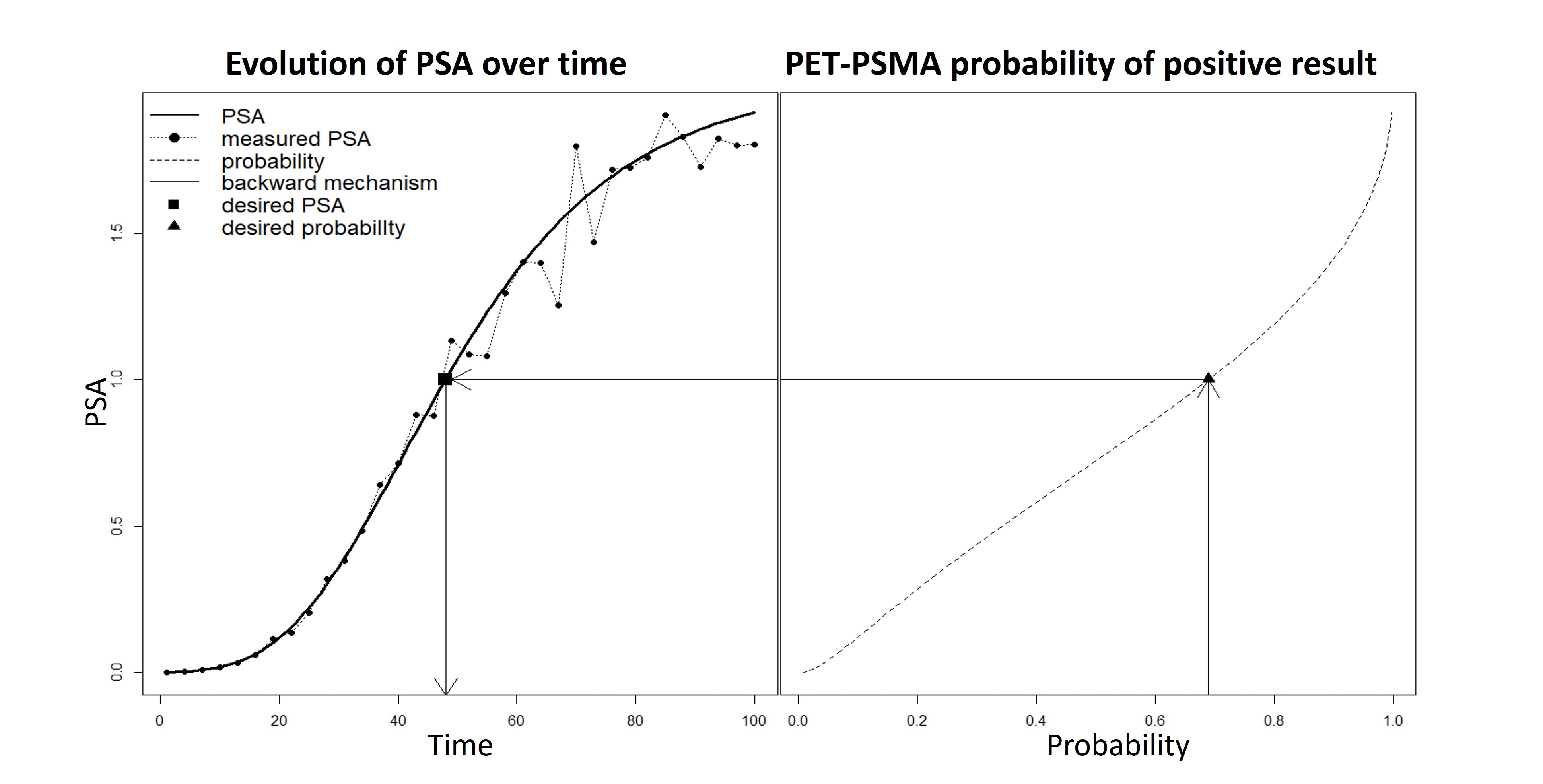}
  \caption{{Joint modeling of $x_i(t)$ and $\pi_i(t)$. The plot shows the relation between time $t$ and PSA evolution $x_i(t)$ (on the left) and between PSA $x_i(t)$  and the probability $\pi_i(t)$ of a positive test (rotated, on the right). After choosing the desired probability, the associated PSA and time can be recovered through the model, following the black arrows.}}
  \label{figoptime}
\end{figure}

In the Bayesian approach, the overarching goal is to compute the posterior density
\begin{equation} \label{eq:densp}
f(\log \mathbf{y}^u, \mathbf{z}^u, \log \mathbf{x}^u,\log \mathbf{x}^o, \boldsymbol{\pi}^u,\boldsymbol{\pi}^o  ,\boldsymbol{\theta}\,|\,\log \mathbf{y}^o,  \mathbf{z}^o),
\end{equation}
based on which all quantities of interest may be computed.
Marginalizing the posterior density in Equation \eqref{eq:densp}, one can obtain, for each fixed time point $t$, the posterior predictive density 
\begin{equation} \label{eq:pred}
f\big(\pi_i(t),\tau_i\,|\,\log \mathbf{y}^o,\mathbf{z}^o \big).
\end{equation}
It should be noted that this can be done since each specific $\tau_i$ is a component of the high dimensional parameter vector $\theta$ and, similarly, for each $t$, $\pi_i(t)$ is a parametric function.
For each $t$, we can then verify whether the following condition is satisfied:
\begin{equation} \label{eq:optime}
\begin{aligned}  
    &\mathcal{P} \big( \pi_i(t) > \pi^* \cap \, t > \tau\,|\,\log \mathbf{y}^o,\mathbf{z}^o\big) = \int_{\pi^*}^{1} \int_{0}^{t} f\big(\pi_i(t),\tau_i\,|\,\log \mathbf{y}^o,\mathbf{z}^o \big) d \tau_i   d  \pi_i(t) \geq \rho ,
    \end{aligned}
\end{equation} 
where $\mathcal{P}$ stands for posterior predictive probability based on the density in Equation \eqref{eq:pred}, $\pi^*$ is a target probability of positive PET-PSMA, and $\rho$ is a posterior assurance probability (say $95\%$), similar to a confidence coefficient.
It should be noted that in Equation \eqref{eq:optime} we require $t$ to be greater than the change point $\tau_i$, whereas $\pi^*$ and $\rho$ are design parameters.
The resulting decisions are strongly dependent on the latter quantities, which should be carefully chosen in advance, in accordance with clinicians.
Default value is $\rho=0.95$, while $\pi^*$ is usually selected with additional ROC analysis or according to clinicians expertise.   
Finally, to select the optimal time $t_i^*$ for the $i$-th patient, 
we can choose the first available time in the future satisfying Equation \eqref{eq:optime}, i.e., the smallest $t$ greater than the largest time in $\mathcal{T}_i$ satisfying Equation \eqref{eq:optime}.
Figure~\ref{figoptime} contains a graphical depiction of the procedure used to obtain the optimal time and how the latent PSA level is connected to the probability of a positive test. In particular, the plot shows the relation between time and PSA evolution, on the left side, and between PSA and the probability of a positive test, on the right side. After choosing the desired probability (on the right-side x-axis), the associated PSA and time can be recovered through the model, following the black arrows, from left to right.

\subsection{Computing the optimal time}
Due to the complexity of the hierarchical joint model, the posterior distribution is defined on high-dimensional data, which prevents us to compute, in closed form, any of the quantities that we may need for inference, as well as normalization constants.
As often done with Bayesian models, we use Markov Chain Monte Carlo (MCMC) \cite{MCMC} algorithms to obtain samples from the density in Equation \eqref{eq:densp} and  Monte Carlo integration \cite{GH} approaches to approximate the posterior quantities of interest, such as posterior expectations and posterior probabilities and, specifically, the integral in Equation \eqref{eq:optime}, which is the main focus of inference.

Let $\log \mathbf{y}^{u,b}, \mathbf{z}^{u,b}, \log \mathbf{x}^{u,b},\log \mathbf{x}^{o,b}, \boldsymbol{\pi}^{u,b},
\boldsymbol{\pi}^{o,b}  ,\boldsymbol{\theta}^{b}$ be
 the $b$-th posterior samples from the associated parameters, where $b = 1,2, \dots , B$, and $B$ is a large number of MCMC iterations.
Using these posterior samples we can approximate the optimal-time $t_i^*$, since samples from the predictive distribution in Equation \eqref{eq:pred} are easily obtainable. From Equations \eqref{eq:logx1}, \eqref{eq:logx} and \eqref{eq:logit} we can see that $\pi_i(t)$, for all $t$, is a deterministic function of the parameters, even if $t$ is not in $\mathcal{T}_i$. 
Therefore, for a  given $t$, and each $b$ sample, we can compute
\begin{equation}
\begin{aligned} 
&\log x_i^b(t)\!=\! \begin{cases}
    \lambda_i^b-\mu_i^bt, 
    & \text{if } t<\tau_i^b, \\
    \log x_i^b (\tau_i^b) \exp(-\gamma_i^b(t-\tau_i^b))+a_i (1- \exp(-\gamma_i^b(t-\tau_i^b))),  
    & \text{otherwise, }
\end{cases}
\end{aligned}
\end{equation}
and
\begin{equation}
\pi_i^b(t) = \frac{\left(x_i^b(t)\right)^{\beta_1^b}e^{\beta_{0,i}^b+\beta_2^bt}}{1+\left(x_i^b(t)\right)^{\beta_1^b}e^{\beta_{0,i}^b+\beta_2^bt}}.
\end{equation}
As a consequence,  the set of samples $\{ \pi_i^b(t), \tau_i^b\}_{b=1}^B$ are from the predictive distribution of Equation \eqref{eq:pred}.
It is easy to show that the integral in Equation \eqref{eq:optime} can be seen as an expectation if we rewrite it as
\begin{equation}\label{eq:mc}
\footnotesize{
\int_{\pi^*}^{1}\!\int_{0}^{t}\!f\big(\pi_i(t),\tau_i\,|\,\log \mathbf{y}^o,\mathbf{z}^o \big) d \tau_i   d  \pi_i(t)  = \int_{0}^{1}\!\int_{0}^{\infty}\!\mathbbm{1}_{[\pi^*,1]}\big(\pi_i(t)\big) \mathbbm{1}_{[0,t )}(\tau_i) f\big(\pi(t),\tau_i\,|\,\log \mathbf{y}^o,\mathbf{z}^o \big)  d \tau_i   d  \pi_i(t),}
\end{equation}
where $\mathbbm{1}_{\cdot}(\cdot)$ is the characteristic function of a set.
Hence, using samples from the predictive distribution, we can approximate the quantity in Equation \eqref{eq:mc} using a standard Monte Carlo integration, leading to the estimator
\begin{equation}\label{eq:mc1}
\begin{split}
 \int_{0}^{1}\!\int_{0}^{\infty}   \mathbbm{1}_{[\pi^*,1]}\big(\pi_i(t)\big) & \mathbbm{1}_{[0,t )}(\tau_i) f\big(\pi(t),\tau_i\,|\,\log \mathbf{y}^o,\mathbf{z}^o \big)  d \tau_i   d  \pi_i(t)  \approx  \frac{\sum_{b=1}^B   \mathbbm{1}_{[\pi^*,1]}\big(\pi_i^b(t)\big) \mathbbm{1}_{[0,t )}(\tau_i^b)}{B}.
\end{split}
\end{equation}
This means that the integral can be approximated by the proportion of posterior samples that are in the set $[\pi_i(t)>\pi^*, \tau_i<t]$.
The approximation in Equation \eqref{eq:mc1} must be computed for a fine grid of time points and the smallest $t \geq \max ( \mathcal{T}_i)$ that satisfies
\begin{equation}
\frac{\sum_{b=1}^B   \mathbbm{1}_{[\pi^*,1]}\big(\pi_i^b(t)\big) \mathbbm{1}_{[0,t )}(\tau_i^b)}{B} \geq \rho
\end{equation}
is the desired optimal time.

\section{Simulations} \label{sec:sim}
\textcolor{black}{ To investigate the model performance
we conduct a simulation study. Due to the complexity of the model, we first examine a single simulated dataset and comment on its results before detailing the full simulation study in Section \ref{sec:simfull}. The simulation is not intended to mimic a real database; rather, it serves as a test to stress the algorithm and analyze its strengths and weaknesses under challenging conditions. The results of a more realistic scenario can be found in Appendix A.2. In the captions of the tables reporting the results, we will use S1 to indicate results from the single simulated dataset discussed in this section, S2 for the results from all datasets, and S3 for the more realistic setting. The results of the model applied to the motivating example, along with a comparison to a competitive approach, are presented in Section \ref{sec:real}.}

\subsection{One simulated dataset (S1)}\label{onedata}
In this section, with a simulated dataset, we want to show that the model, and especially the change point values $\tau_i$, can be estimated using the MCMC algorithm.  We simulate a scenario where $m=80$ patients are undergoing surgery at time 0 and then followed up for several months.
For each patient, we simulate the number of elements of $\mathcal{T}_{y,i}$ from the distribution $U_d(5,8)$, where $U_d(\,\cdot\,,\,\cdot\,)$ indicates a uniform distribution over the integers between the two arguments (included). The times associated with the measurements are sampled randomly without replacement from the set $\{1,2,\dots, 25\}$, while the numbers of time points of the PET-PSMA exams are from the $U_d(3,5)$ and the times 
are randomly sampled without replacement from the set $\{26,27,\dots , 38\}$.  
For each $i$, we assume $\min\{{\mathcal{T}_{z,i}}\}>\max\{{\mathcal{T}_{y,i}}\}$, meaning that the exams are always performed after the last  PSA measurement and we have a small set of measurements for each patient. Let $\mathcal{T}_{y,i}=\{t_{i,j}\}_{j=1}^{n_i}$ be the set of ordered points:
given these values, to simulate $\tau_i$ we sample from $U(t_{i,3},t_{i,(n_i-2)})$, so that
$\tau_i$ is in the middle of the temporal window. 
We simulate data in this way to create a challenging situation, where the data points are very few, and the PET-PSMA exams $\{z_i(t)\}_i$ are all performed much later than the last PSA measurement $\{y_i(t)\}_i$. This is done to highlight better how the method can be used to predict the PET-PSMA results for future time $t$, for which we have not observed the PSA level, and when data information is poor.
For each patient $i$, we simulate 9 dichotomous variables $\{C_{ij}\}_{j=1}^{9}$, and a final variable $C_{i10}=\lfloor \hat C_{i10} \rfloor$, with $\hat C_{i10}\stackrel{i.i.d.}{\sim}\mathcal{N}(75,7)$, used  to describe the patient age.
We then assume 
$\mathbf{C}_{i, \boldsymbol{\mu}} = (1,C_{i1},C_{i2},C_{i3},C_{i4},C_{i5} ) $ and $C_{i,\boldsymbol{\beta}}=(1,C_{i6},C_{i7},C_{i8},C_{i9},C_{i10})$. The remaining parameters are 
\begin{equation}
\begin{aligned}  
 & \boldsymbol{\alpha}_{\boldsymbol{\mu}} =  \left(
 \begin{array}{c}
1\\ 0.1\\0.3\\0.5\\0.2\\0.1
\end{array} 
\right), \,\,\,\,
\boldsymbol{\alpha}_{\boldsymbol{\gamma}} =  \left(
  \begin{array}{c}
  -1\\ -0.01\\-0.01\\-0.01\\-0.01\\-0.01
  \end{array} 
  \right), \,\,\,\, \boldsymbol{\alpha}_{\boldsymbol{\beta}} =  \left(
    \begin{array}{c}
    1\\  1\\1 \\0.5\\-0.5\\-0.5
    \end{array} 
    \right),\,\,\,\,
  \left(
    \begin{array}{c}
    \beta_1 \\\beta_2  \\ \psi_{a} \\ \omega_{\boldsymbol{\mu}}\\ \omega_{\boldsymbol{\gamma}}  \\ 
    \omega_{a} \\
    a_{\sigma^2} \\
    b_{\sigma^2} \\
    \end{array} 
    \right) =  \left(
  \begin{array}{c}
  4\\  0.5\\5.7 \\0.1\\0.1\\1 \\ 3 \\ 5\\
  \end{array} 
  \right).
  \end{aligned}
\end{equation}
Note that most of the parameters chosen for the simulation are randomly selected, and there is no intention to be realistic.
A Gaussian prior $\mathcal{N}(0,100)$ is used on 
\begin{equation}
    \lambda_i,\, \log \omega_{\mu},\, \log \omega_{\gamma},\, \psi_a,\, \log \frac{b_{\sigma^2}}{(a_{\sigma^2}-1)},\, \log{\frac{b_{\sigma^2}^2}{(a_{\sigma^2}-1)^2(a_{\sigma^2}-2)}},
\end{equation}
and all regression coefficients, where $b_{\sigma^2}/ (a_{\sigma^2}-1)$ and ${b_{\sigma^2}^2}/\left({(a_{\sigma^2}-1)^2(a_{\sigma^2}-2)}\right)$ are respectively mean and variance of the parameter $\sigma_i^2$, while $\omega_a\sim IG(1,1)$. \textcolor{black}{Given the range of value that we expect in the real data application, and the values that we use to simulate the data, the normal priors can be considered weakly informative, as they put most of their mass on likely values without being too informative, and standard choice for the parameters of the inverse gamma are chosen.}

The prior of $\tau_i$ is a patient-specific mixed-type distribution:
 $\tau_i$ can assume value in  $\{t_{i1},[t_{i2},t_{iJ_{i-1}}],t_{iJ_i}\}$, where  $\{t_{i1},t_{i2},...,t_{iJ_i}\} \equiv \mathcal{T}_{y,i}$, with a probability mass of 1/3 on $t_{i1}$ and $t_{iJ_i}$, and a Uniform distribution in $[t_{i2},t_{iJ_{i-1}}]$. The reasoning behind this choice is that a change point in $[0, t_{i2})$  is not identifiable, since only the observation collected at time $t_{i1}$ is available to estimate the decreasing phase. The same applies in the time window $(t_{iJ_{i-1}},+\infty)$ where we could gain information only from $t_{iJ_{i}}$ to estimate the log Gompertz coefficients. Since the change point is one of the most important parameters for the final identification of the optimal time, we prefer to have a parameter that is always identifiable. 
We run our algorithm for $150000$ iterations, with burn-in equal to $100000$ and thinning parameter equal to 10.
The reason to discard such a long burn-in sequence is that the model is 
fairly complex,
and we want to make sure that a reasonable hope for convergence holds.
In addition, the model contains hyper-parameters over which extremely flat
prior distributions have been put.
As MCMC diagnostics we use the $\hat R$ (see \cite{WAIC}), \textcolor{black}{an index specifically designed to check mixing and stationarity of the posterior chains, by looking at their inter- and intra-variabilities.} 
The algorithm is a Metropolis-within-Gibbs MCMC with the adaptive Metropolis steps proposed in \cite{andrieu2008tutorial}, and a P\'olya-Gamma update \cite{polson2013bayesian}
for $(\beta_1,\beta_2,\alpha_{\beta})$.
\textcolor{black}{The model is implemented in \texttt{R}, and for each dataset approximately seven hours are needed to obtain posterior samples.}
 
%

\textcolor{black}{In Table \ref{table_simuation_ind}, for each individual parameter, we show the posterior \(95\%\) credible interval (CI) and evaluate the percentage of parameters correctly estimated (for which the true values used for simulation are inside the \(95\%\) CI). From Table \ref{table_simuation_ind}, we see that the percentages calculated for more than 80 patients are close to \(95\%\), with the median \(\hat{R}\) being very close to the optimal value of 1 \cite{WAIC}. }
\textcolor{black}{
In terms of global parameters, see Table \ref{table_simuation_glob},  all CIs of the global parameters contain the true values used to simulate the data,  with   good $\hat{R}$ values  but the CIs of the logistic parameters tend to be large, particularly for  $\alpha_{\beta}[1]$. It should be noted that the true value of $\beta_2$ coincides with the CI lower limit.}

\begin{table}[t]
\centering
\caption{S1 - Individual parameters. The first column of the table shows the percentage of correctly estimated parameters across the 80 individuals. The second column contains the empirical quantiles at levels \([2.5 \%, 97.5 \%]\) and the sample mean (the value in the center) of the distribution of \(\hat{R}\) across the 80 individuals. } 
\label{table_simuation_ind}  
\begin{tabular}{c||cccc}
\hline
 Parameter & Percentage  & $ CI_{\hat R}$   \\ \hline
 $\log(\lambda_i)$ & 95\% & [1.00 1.00 1.01] \\
 $\tau_i$ & 95\% & [1.00 1.00 1.00]\\
  $\log(\gamma_i)$ & 91\% & [1.00 1.00 1.00]\\
 $\log(\mu_i)$ & 95\% &  [1.00 1.00 1.01]\\
 $\log(a_i)$ &100\% & [1.00 1.00 1.00] \\
 $\sigma^2_i$ & 96\% & [1.00 1.00 1.00]\\
  \hline
\end{tabular}
\end{table}

\subsection{Full simulation study (S2)}\label{sec:simfull}
The full simulation study comprises four scenarios: scenario 1 and scenario 2 differ only in the noise value (determined by parameters $a_{\sigma^2}$ and $b_{\sigma^2}$), while the third scenario differs in most of the parameters. Scenario 4 closely resembles scenario 1 but is composed of fewer data points.
Specifically, in scenarios 1-2-3, the datasets consist of 80 patients, with each patient having between 8 and 15 PSA measurements and between 3 and 5 PET-PSMA examinations. In scenario 4, the dataset is composed of 80 patients, with each patient having between 6 and 10 PSA measurements, and between 1 and 3 PET-PSMA examinations. The underlying parameters used to simulate the datasets are summarized in Table \ref{scenarios_table}.
For each of the four scenarios, we simulate 100 datasets and run the MCMC procedure to obtain samples from the posterior. The results are summarized in Tables \ref{tab:simulation_full2} and \ref{tab:simulation_full1}-\ref{tab:simulation_full1_2}, reporting the percentage of individual and global parameters correctly estimated (the true values are within the posterior $95\%$ interval).
Additionally, the tables include the quantiles at levels $2.5\%$ and $97.5\%$, the mean of the interval widths, and $\hat{R}$, computed across the 100 datasets of each scenario.

\textcolor{black}{
Except for the variance of the random effects, the model performance in predicting PSA levels tends to be better than in the logistic component. If the number of collected PET-PSMA measurements decreases, although the percentage of correctly estimated parameters remains relatively high, the amplitude of the posterior credible interval increases, resulting in reduced precision in estimating the parameters for the logistic model.   This highlights how the quality of estimation and the reliability of the results strongly depend on the number of available data points for each patient.
In particular, the quality of the estimates for the logistic regression parameters depends heavily on the scenario and how much information that scenario provides about the specific parameters, with percentages going as low as 5\% (especially, for Scenario 3 which turns out to reflect a very challenging case) and, at times, percentages close to 95\% being associated with large CI widths. Moreover, the simulation study highlights that higher values of the noise term lead to a decrease in the model efficiency in estimating the growth coefficients \(\mu_i\) and \(\gamma_i\), which can be seen by the interval widths in Table \ref{tab:simulation_full2}. However, the overall performance of the model is still satisfactory. These results suggest that, for practical application in a clinical setting, a large dataset is necessary to accurately learn the relationship between PSA levels and test outcomes. Once this relationship is established, it can be effectively used to predict outcomes for future patients, even in cases where limited data and information are available.}

\textcolor{black}{We further assess the model under a more realistic range of parameters, with results presented in Appendix A.2. In this setting, we simulate 100 datasets using the mean parameter estimates obtained from the real case study (detailed below, in Section \ref{sec:real}) while assuming a similar dataset composition. Under this approach, all diagnostics used to evaluate the model ability to retrieve the true values used in data simulation show improvements, particularly for those parameters that performed poorly in simulations S1 and S2. These results serve as a sensitivity analysis for the case study, confirming the reliability of the model and algorithm.
}

\begin{table}[t]
    \centering
    \caption{S2 -  Individual parameters. For each scenario (columns) and parameter (rows), the first row reports the percentage of correctly estimated parameters. The second row shows the empirical quantiles at levels \([2.5 \%, 97.5 \%]\) and the sample mean (the value in the center) of the distribution of the confidence interval (CI) widths. The third row displays the empirical quantiles at levels \([2.5 \%, 97.5 \%]\) and the sample mean (the value in the center) of \(\hat{R}\). All components of the table are computed across the 80 individuals and 100 datasets.}
    \setlength{\tabcolsep}{2pt} 
    \begin{tabular}{rc||cccc}
    \hline
      \
    {Parameter}& & Scenario 1 & Scenario 2 & Scenario 3 & Scenario 4 \\
    \hline
   &\% & 95$\%$  &  93$\%$ &  94$\%$  &  95$\%$ \\
   $\log(\lambda_i)$&  $ CI_{W} $ & [3.11 5.51 9.98]  &  [4.39 7.44 12.76] &  [4.11 8.96 18.66] &  [3.11 5.52 9.99]\\
   &   $ CI_{\hat R} $ & [1.00  1.02  1.10]  &  [1.00  1.03  1.19] &  [1.00  1.04  1.30] &  [1.00  1.02  1.10]\\
    \hline
    & \%          & 98$\%$   &  96$\%$ &  95$\%$ &  98$\%$\\
    $\tau_i$ & $CI_{W} $          &  [0.26 1.16 3.52]  &   [0.40 1.56  4.58] &  [0.34 2.25  6.98] &  [0.26 1.16 3.53]\\
    &  $CI_{\hat R} $ & [1.00  1.01  1.06]  &  [1.00  1.02  1.11] &  [1.00  1.01  1.10] &  [1.00  1.01  1.06]\\
       \hline
    & \% & 98$\%$  &  90$\%$ &  96$\%$  &  98$\%$ \\
    $\log(\gamma_i)$  &$ CI_{W}$  & [0.10 0.26 0.44] &  [0.00 0.28  0.52] &   [0.01 0.43  1.93] &  [0.10 0.26 0.43]\\
  & $CI_{\hat R} $ & [1.00  1.01  1.08]  &  [1.00  1.03  1.16] &  [1.00  1.03  1.24] &  [1.00  1.01  1.08]\\
   \hline
    & \%     & 95$\%$  &  90$\%$  &  93$\%$  &  95$\%$ \\
    $\log(\mu_i)$  & $CI_{W} $     & [0.30 0.72 1.45] &   [0.33 0.85  1.67] &  [0.4 1.43  3.91] &   [0.30 0.72 1.46]\\
    &  $CI_{\hat R} $ & [1.00  1.02  1.14]  &  [1.00  1.05  1.33] &  [1.00  1.01  1.07] &  [1.00  1.02  1.13]\\

    \hline
    &   \%      & 94$\%$&  93$\%$  &  97$\%$  &  94$\%$ \\
    $\log(a_i)$ &   $CI_{W} $       &  [2.39 4.04 6.36] &  [2.72 4.66  8.03] &   [2.56 4.52  7.10] & [2.40 4.04 6.36]\\
    &    $CI_{\hat R} $ & [1.01  1.04  1.22]  &  [1.00  1.08  1.52] &  [1.00  1.02  1.10] &  [1.00  1.04  1.21]\\
    \hline
    &   \%      & 88$\%$   &  91$\%$  &  94$\%$  &  88$\%$ \\
  $\sigma_i^2$   &     $ CI_{W}$      & [0.94 1.62 2.89]  &   [1.39 2.23 3.96] &   [1.07 1.95  4.04] &   [0.94 1.62 2.90]\\
    &      $ CI_{\hat R}$ & [1.00  1.01  1.08]  &  [1.00  1.02  1.11] &  [1.00  1.00  1.03] &  [1.00  1.01  1.08]\\

   \hline
   
    \end{tabular}\label{tab:simulation_full2}
\end{table}

\section{Application to clinical data} \label{sec:real}
\begin{figure}[h!]
  \centering
  \includegraphics[scale=0.55]{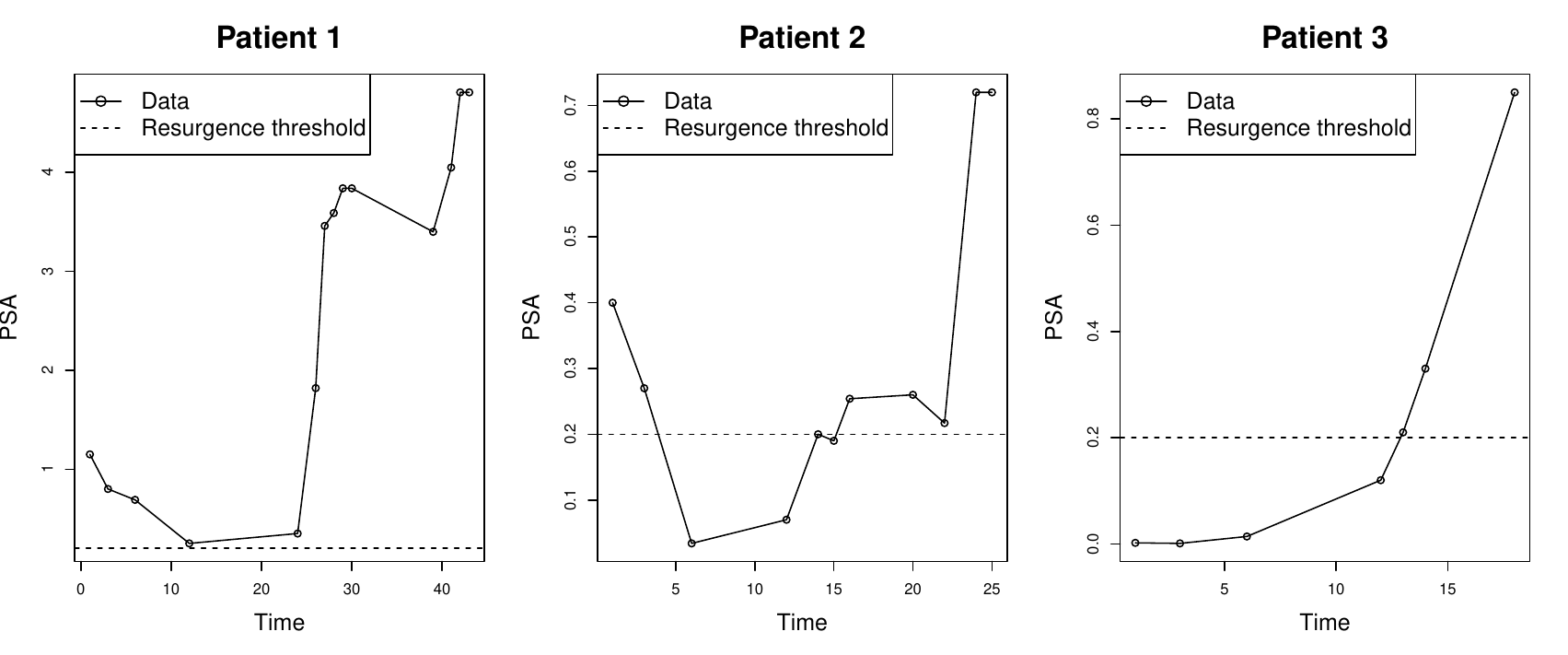}
  \caption{PSA measurements collected on three patients. Patients 1 and 2 are of type BCP but with quite different PSA levels, while 3 is of type BCR.}
  \label{data}
\end{figure}

The database is built on clinical practice in the San Luigi Hospital in Torino. In prostate cancer follow-up setting post radical prostatectomy (RP), PSA is the main source of information to base clinical decisions 
regarding interventions such as PET-PSMA exam. 
We have $n = 187$ patients and, for each of them, we have several demographic and clinical variables, as well as the PSA measurements and PET-PSMA results taken at different times after RP. The number of time points in $\mathcal{T}_{y,i}$ ranges from  4 to 17, while the ones in  $\mathcal{T}_{z,i}$ from 1 to 4, and the follow-up period is between 4 and 280 months.
We can distinguish between patients who, after prostatectomy, show relatively low values of PSA after RP, but may be eventually subject to a biochemical relapse at change point (BCR patients), and Biochemical-persistence patients (BCP patients), who present persistent benign/malignant residual tissue after surgery (conventionally signaled by PSA $>$ 0.2) and are usually treated with therapy to inhibit cancer growth - for them, PSA levels may even initially decrease until a new increase occurs at the change point. The presence of a mixed population of BCR and BCP patients and the possible administration of different therapies make the inference task a very difficult exercise. 
To present the possible data paths, Figure \ref{data} shows data collected for some anonymous patients, that we refer to from now on as patients 1, 2, and 3. The shape of the data for these patients is quite different (patients 1 and 2 are BCP, while patient 3 is BCR), but also similar shapes can correspond to quite different PSA values (patient 1 and patient 2). 
Before surgery, each patient is assigned to a different category according to the status ($S_i$), but some more information is collected at surgery time; in particular, pathological stadiation according to TNM classification, and the clinical status of prostate margins. Regarding tumor status $P^T_i$, the clinicians use four different categories:
$T1$ (clinically unapparent tumor), $T2$ (tumor confined within prostate),
$T3$ (tumor extends through the prostate capsule), $T4$ (tumor is fixed or invades adjacent structures). For the sake of simplicity,
here the 4 tumor categories have been reduced to two:
$P^T_i=0$ if $T_1$ and $T_2$ and $P^T_i=1$ otherwise.
The lymph nodes implication is evaluated through a binary variable ($P_i^N$), while metastases are always present and, thus, not relevant.
Moreover, the clinical stage of the tumor is usually represented using the Gleason Score \cite{GLEASON-SCORE}, and we reduced the original nine categories to two:
$S_i=0$ if Gleason Score is less than 6 or if it is equal to 3+4, $S_i=1$ otherwise.
Finally, prostate resection margins ($P_i^R$) are evaluated after the surgery.
\newline \indent  According to clinical evaluation, each patient with BCR and/or BCP can then be administered four different therapies: adjuvant or salvage androgen deprivation hormone therapy ($\text{OA}_i$ or $\text{OS}_i$), adjuvant or salvage radiotherapy ($\text{RA}_i$ or $\text{RS}_i$). Some patients also underwent regional lymphadenectomy ($\text{L}_i$). Adjuvant therapies are administered within 6 months from surgery, while salvage ones are performed after six months on from surgery. In addition, for each patient, we know the age $A_i$. All these data are used as clinical and demographic covariates to gain information useful for our model. \textcolor{black}{A descriptive table of the analyzed cohort of patients can be found in Appendix (see Table \ref{summary-cov}).}

\subsection{Results}
\begin{figure}[h!]
  \centering
 \includegraphics[scale=0.3]{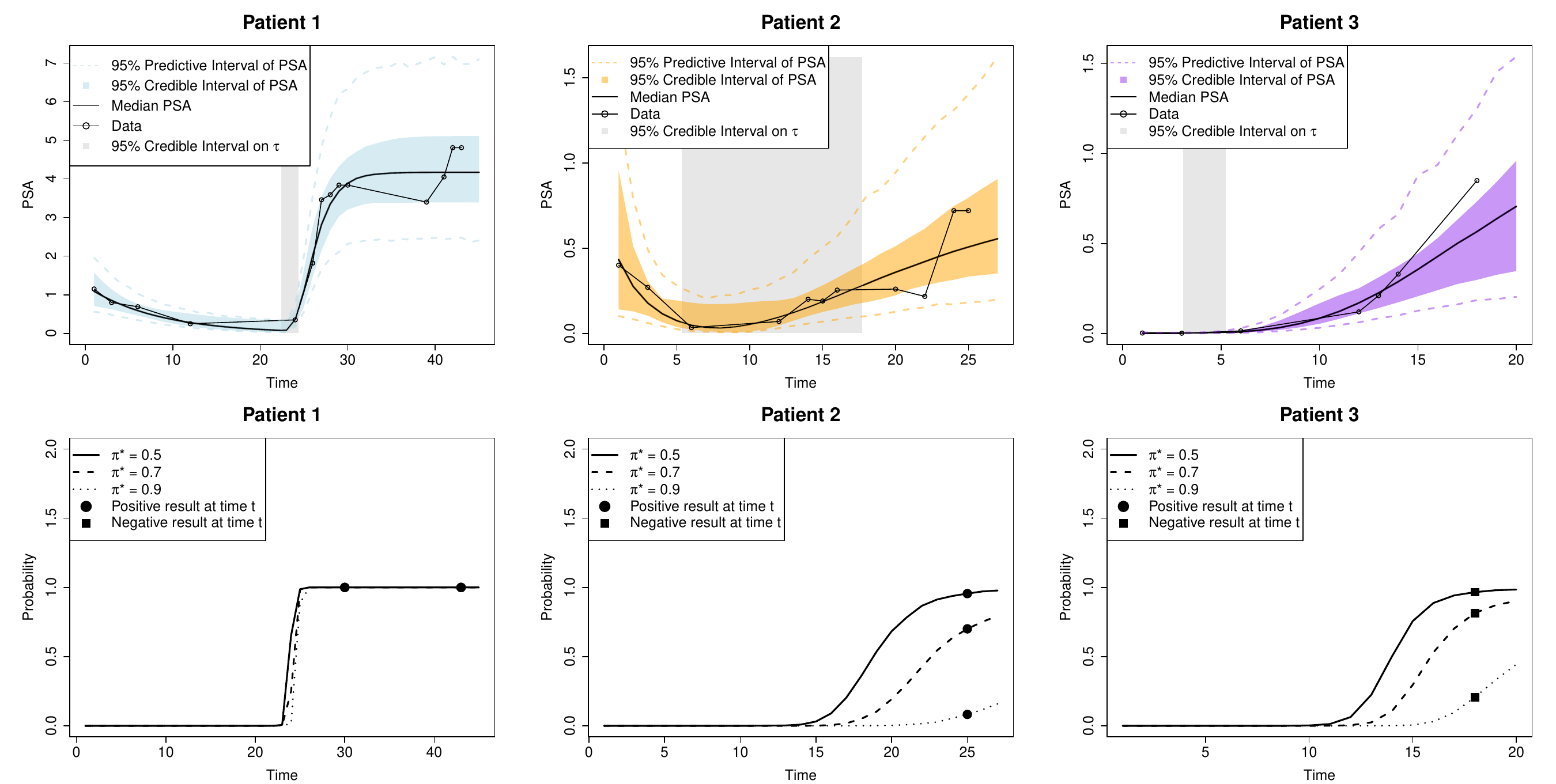}
  \caption{PSA growth curves (on top) and probability curves (on bottom) for the three patients introduced in Figure \ref{data}. Patients 1 and 2 are BCP but with quite different PSA levels, while 3 is BCR.}
    \label{trj}
\end{figure}

\begin{figure}[h!]
  \centering
 \includegraphics[scale=0.3]{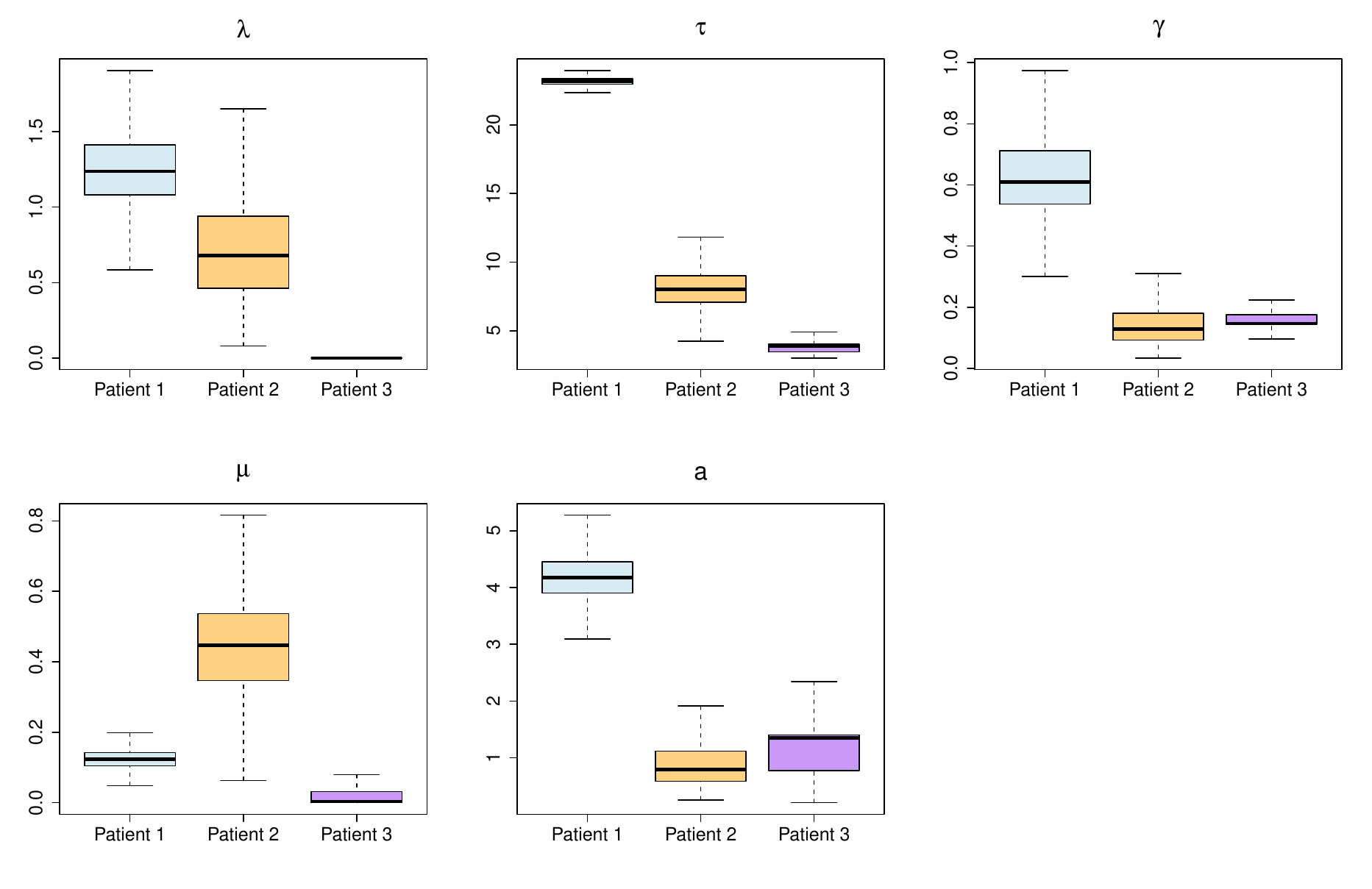}
  \caption{Individual parameters posterior distributions for the three patients introduced in Figure \ref{data}.}
    \label{phip}
\end{figure}

In particular, we assume that $\mu_i$, $\gamma_i$, $\sigma_i^2$ and $a_i$ are random effects, with $\psi_{i,a}=\psi_{a}$, $a_{i,\sigma^2}$=$a_{\sigma^2}$, and $b_{i,\sigma^2}$=$b_{\sigma^2}$ (constant for each patient), while the means of both $\mu_i$, $\gamma_i$ depend on covariates, to gain more information from the available data. However, the observed difference between the reported categories of patients, namely BCR and  BCP patients, suggests that the parameters $\lambda_i$ do not come from a common population. Instead of constructing a bimodal random effect which is out of our scope, we take each $\lambda_i$ to be an individual parameter.
We estimated three models with different combinations of covariates  $\mathbf{C}_{\mu}$,  $\mathbf{C}_{\gamma}$, $\mathbf{C}_{\pi}$, all suggested by clinical evidence, and among them, we selected the best model using the Watanabe–Akaike information criterion (WAIC) \cite{WAIC}, see Table \ref{table_simuation_global} in Appendix. The results we describe refer to the best configuration selected, namely Model 2.
In Tables \ref{table_mu}, \ref{table_gamma}, \ref{table_logit}, and \ref{table_global} we show the global parameter estimates for the best model, 
with relative $95\%$ CIs.
\paragraph{Covariates interpretation}
 Interpretation of coefficients can be difficult and misleading, particularly when referring to therapies. These are the usual difficulties in interpreting causality with observational and clinical practice databases rather than clinical trials.
From a clinical perspective, one could expect therapies to decrease the PSA level before change point $\tau$ and to reduce the growth speed after it.
However, the analysis is not targeting therapeutic efficacy. Therapies are used here as indicators of the severity of the disease rather than for estimating their effect. For this reason, we report the logistic analyses used to determine the relationship between administrations of therapies and baseline covariates (see Table \ref{table_logit_terapie}), where it is easy to notice that,
as expected, patients with the worst clinical situation at the surgery time have higher probabilities of receiving one or more of the analyzed therapies. 
On the other hand, baseline covariates have relatively simple interpretations. Following the theory, PSA decreasing level turns out to be almost zero for patients that do not receive therapies (see $\alpha_{\mu}[1]$). Gleason score is a significant factor, as higher levels determine an increase in PSA values. In particular, the standard deviation of decreasing coefficient $\omega_{\mu}$ and increasing coefficient $\omega_{\gamma}$ are quite different. The high level of the former reflects the heterogeneity of the dataset: PSA levels right after surgery and preceding the changing points are different for BCR and BCP. Finally, the probability of a positive PET-PSMA result at time $t$ strongly depends on the PSA level at $t$ (as was to be expected), on the Gleason score, and on lymph nodes implication. In particular, as the PSA level increases, also the probability increases.
\paragraph{Graphical results} In Figure \ref{trj}  we show examples of the model outputs. In the first row, the fitted curves for the selected patients (first row) are shown, where the grey solid regions are used to show the $95\%$ CI of $\tau_i$. In the second row, the approximation on the right-hand side of Equation \eqref{eq:mc1} is computed using thresholds 0.5, 0.7, and 0.9: as expected, the higher the threshold, the lower the probability. Filled circles and squares are used to indicate the true and negative outcomes of the test, respectively.
The posterior distributions of individual parameters $\lambda_i, \ \tau_i,\ \gamma_i,\ \mu_i,\ a_i$ are reported in Figure \ref{phip}. The posterior distributions  $\lambda_i$ show differences that reflect the difference between BCP and BCR patients. In particular, it is easy to see that we obtain different fits for patients 1 and 2 of type BCP, in comparison with BCR patients 3. 
\newline \indent
\textcolor{black}{ 
  It should be noted that, in this case study, data were collected in an observational setup rather than in a clinical trial. Therefore, the timing of measurements was not pre-specified but was influenced by the clinician's decisions and the severity of the illness itself, which is a potential source of bias. We recognize that this is a shortcoming of our analysis. However, modeling the timing of the tests, which would require a point-process modeling approach, would drastically increase the complexity of the model. We do not believe that there is enough information in the data to estimate the parameters of this additional component accurately, but it would be part of future analyses. }

\begin{table}[t]
\centering
\caption{Real Data Application - Global parameters describing the PSA level of the proposed model before the change point. The table shows the limits of the 95\% CIs (2.5\% and 97.5\%), the posterior mean (Mean), and the value of the $\hat{R}$ statistic.
 Positive coefficients are associated with a lower level of PSA. }
\begin{tabular}{c||cccc}
\hline
Parameter &2.5\%  & Mean & 97.5\% & $\hat R$   \\ \hline
$\alpha_{\mu}[1]$ - Intercept            & -10.37 & -7.76  & -5.81 & 1.13 \\
$\alpha_{\mu}[2]$ - Ormono adj           & -0.20  & 1.84   & 3.84  & 1.02\\
$\alpha_{\mu}[3]$ - Ormono salvage           & -5.67  & -3.18  & -0.49 & 1.05 \\
$\alpha_{\mu}[4]$ - Radio adj            & -1.66  & 0.58   & 2.85 & 1.00 \\
$\alpha_{\mu}[5]$ - Radio salvage            & -0.53  & 1.46   & 3.36 & 1.00 \\
$\alpha_{\mu}[6]$ -  Lymphadenectomy       & -1.30  & 1.13   & 4.25   & 1.20 \\
\hline
\end{tabular}
\label{table_mu}
\end{table}

\begin{table}[t]
\centering
\caption{Real Data Application - Global parameters describing the PSA level of the proposed model after the change point. The table shows the limits of the 95\% CIs (2.5\% and 97.5\%), the posterior mean (Mean), and the value of the $\hat{R}$ statistic.
Positive coefficients are associated with a lower level of PSA.}
\begin{tabular}{c||cccc}
\hline
Parameter &2.5\%  & Mean & 97.5\%   & $\hat R$\\ \hline 
$\alpha_{\gamma}[1]$ - intercept         & -2.99  & -2.58  & -2.18 &1.06\\ 
$\alpha_{\gamma}[2] - P^R$               & -0.14  & 0.20    & 0.56 &1.03\\
$\alpha_{\gamma}[3] - P^T$               & -0.29  & 0.08   & 0.44 & 1.00\\
$\alpha_{\gamma}[4] - P^N$               & -0.78  & -0.17  &  0.47 & 1.00\\
$\alpha_{\gamma}[5] - S $              & 0.32   & 0.64   & 1.03 & 1.07\\ 
$\alpha_{\gamma}[6]$ - Age               &-0.25   & -0.04  & 0.14 & 1.03\\
$\alpha_{\gamma}[7]$ - Ormono adj        & 0.18   &  0.71  & 1.28 & 1.00\\
$\alpha_{\gamma}[8]$ - Ormono slv        & 0.26   & 0.73   & 1.28 & 1.03\\
$\alpha_{\gamma}[9]$ - Radio adj         & -1.69  &-1.13   & 0.547 & 1.00\\
$\alpha_{\gamma}[10]$ - Radio Slv         & -0.65  &-0.24   & 0.21 & 1.00\\
$\alpha_{\gamma}[11]$ -  Lymphadenectomy        & -0.41  & 0.03  & 0.49 &1.05  \\
\hline
\end{tabular}
\label{table_gamma}
\end{table}

\begin{table}[t]
\centering
\caption{Real Data Application - Global parameters describing the logistic transformation of the proposed model. The table shows the limits of the 95\% CIs (2.5\% and 97.5\%), the posterior mean (Mean), and the value of the $\hat{R}$ statistic.
Positive coefficients are associated with higher probabilities of positive PET-PSMA examinations.}
\begin{tabular}{c||cccc}
\hline
 Parameter &2.5\%  & Mean & 97.5\% & $\hat R$     \\ \hline 
$\alpha_{\beta}[1]$  - Intercept     & -0.59 & 0.84  & 2.34 &1.02 \\
$\alpha_{\beta}[2] - P^R $           & -0.11 & 0.68  & 1.62 &1.01 \\
$\alpha_{\beta}[3]  - P^T  $         & -0.84 & 0.22  & 0.85 &1.00 \\
$\alpha_{\beta}[4]  - P^N $          & 0.05  & 0.88  & 2.56 &1.00 \\
$\alpha_{\beta}[5]  - S $          & 0.05  & 1.28  & 2.56  &1.00\\
$\alpha_{\beta}[6]$ - Age            &-0.52  & -0.14 & 0.24 &1.00 \\ 
$\beta_1$                            & 1.64  & 2.56  & 3.68  & 1.00\\
$\beta_2$                            &-0.01  & 0.00  & 0.01  &1.02\\
\hline
\end{tabular}
\label{table_logit}
\end{table}

\begin{table}[H]
\centering
\caption{Real Data Application -   Random effects parameters of the proposed model. The table shows the limits of the 95\% CIs (2.5\% and 97.5\%), the posterior mean (Mean), and the value of the $\hat{R}$ statistic.}
\begin{tabular}{c||cccc}
\hline
 Parameter &2.5\%  & Mean & 97.5\%   & $\hat R$   \\ \hline
$\omega_{\mu}$                       & 2.53  & 3.14  & 3.91 & 1.00 \\
$\omega_{\gamma}$                    & 0.63  & 0.75  & 0.90  & 1.05\\ 
$\Psi_a$                             &-0.41  & -0.26 & -0.13 &1.01 \\
$\omega_a$                           & 0.57   & 0.66  & 0.76 &1.01\\
$a_{\sigma^2}$                       & 2.00  & 2.01  & 2.02  &1.00\\
$b_{\sigma^2}$                       & 0.18  & 0.22  & 0.27 &1.01\\
\hline
\end{tabular}
\label{table_global}
\end{table}

\begin{table}[H]
\centering
\caption{Real Data Application -  Coefficients of the logistic regression used to determine the relationship between administrations of therapies and baseline covariates. 
Coefficients in bold are associated to p-values smaller than 0.05.
The following abbreviations are used: OA for hormone adjuvant, OS for hormone salvage, RA  for radio adjuvant, RS: radio salvage, L  for lymphadenectomy, PSA-s  for PSA level at surgery.}
\setlength\tabcolsep{1.5mm}
\begin{tabular}{c||ccccccc}
\hline
Therapy         & Intercept     & $P^R$         & $P^T$ & $P^N$         & S             & L             & PSA-s\\ \hline 
OA &\textbf{-3.22} & -0.02         & -0.46 & \textbf{1.92} &  1.20         & 0.46          & \textbf{1.54}  \\
OS  &\textbf{-1.57} & -0.08         & -0.95 & -0.51         &  0.90         &  0.54         & -1.39          \\
RA  &\textbf{-2.94} & \textbf{2.21} & -0.57 & 0.92          & 0.35          & -0.30         & -0.25          \\
RS  &\textbf{-3.14} & 0.52          & 1.02  & -0.70         & 1.01          &  0.23         & -0.79          \\
L &-0.03          & -0.26         & 0.84  & N.A.          & \textbf{1.12} &\textbf{-0.60} & 0.25           \\
\hline
\end{tabular}
\label{table_logit_terapie}
\end{table}

\subsection{Comparison of the joint model to the logistic model} \label{sectionROC}
\begin{figure}[h!]
\centering
 \includegraphics[scale=0.7]{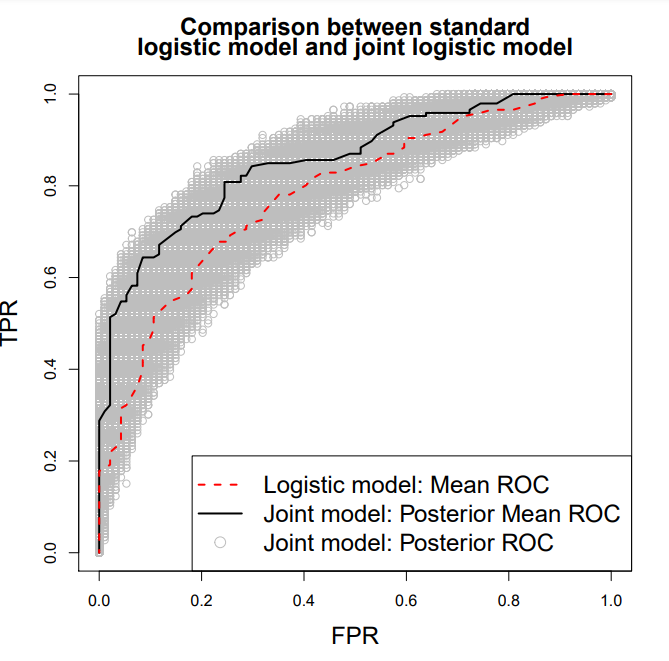}
  \caption{Real Data Application - ROC curves for the results of the simple logistic model and joint logistic model. The AUC for the simple logistic ROC is 0.79, the AUC for the mean joint model ROC is 0.86, while the posterior 95\% distribution of the AUC index ranges between 0.78 and 0.86. Results are obtained with \texttt{R} package \texttt{pROC} \cite{rocpck}.}
    \label{roc}
\end{figure}

We compare the performance of our joint model to the closest method existing in the literature \cite{luiting2020optimal}, that can be considered the current standard of care. In particular, a logistic regression model is fit with the \texttt{R} function \texttt{glm} \cite{stats} ({following the idea proposed in \cite{luiting2020optimal}}), including the same baseline covariates as our joint-logistic model and the observed PSA level. 
In this logistic model, which we compare to our joint model, the response is $1$ if the PET-PSMA gives a positive result (i.e., the location of the disease is identified) and $0$ if not.
We apply to the San Luigi Hospital dataset presented in Section \ref{sec:real} both methods.
Overall, the results obtained from this logistic analysis (see Table \ref{logitTABLE}) are largely consistent with those from our method. Moreover, the significant covariates are also overall consistent with the reference paper \cite{luiting2020optimal}, based on a different but similar dataset. Specifically, we found that tumor status and log-PSA level have strong and significant effects on the probability of positive PET-PSMA results, while resection margin and time have weaker effects. Notably, all of these covariates have positive coefficients, indicating that higher values increase the likelihood of positive PET-PSMA results. To formally compare the two fitted models, we use them to predict the PET-PSMA binary outputs, on the dataset under analysis, and we evaluate the relative Receiver Operating Characteristic curves (ROCs), shown in Figure \ref{roc}, computing their areas under the curve (AUCs). 
\textcolor{black}{Figure \ref{roc} shows the ROC curves for the logistic and the joint model, computed using the \texttt{auc} function from the \texttt{pROC} package \cite{rocpck}. The AUC for the simple logistic model (red line) is 0.79, while the AUC for the mean joint model (black line) is 0.86, with \(95\%\) of the posterior AUC values ranging between 0.78 and 0.85, and a median of 0.81. As a result, we can see that the AUC of the simple logistic model is close to the lower quantile obtained using our approach. This suggests that the use of the latent PSA level, which is one of the main points of our proposal, instead of the measured level used in the competing model
\cite{luiting2020optimal}, tends to yield better results.  From the mean ROC curves, we can select the optimal $\pi^*$ probability, namely the one corresponding to the higher trade-off between sensitivity and specificity, to be used for predicting new data; it is selected as the point on the ROC closest to coordinates $(0,1)$, which corresponds to $\pi^*=0.55$.} Moreover, one big advantage of our joint model is that it enables the prediction of probabilities of positive results for future times, for which the PSA level may still be unknown. For example, this prediction of the PSA level and the related probability of positive success can still impact the treatment strategy adopted in the meantime, thus improving the patient
benefit. 
\newline \indent In addition to the logistic regression presented in \cite{luiting2020optimal} and analyzed in this Section, a common procedure currently used by clinicians is the evaluation of PSA-doubling time (PSA-DT) and PSA-velocity, based on the last couple of PSA measurements. In literature, PSA-DT higher than six months is significantly associated with a high probability of positive PET-PSMA results, but no clear and well-defined guidance is available, resulting in subjective decisions of time to examination, which depend on the clinician's belief and experiences. For this reason, a comparison of the performances of our proposed joint model and this methodology cannot be performed.
\begin{table}[t]
\centering
\caption{Real Data Application -   The Table describes the results of the logistic model based on the proposal of \cite{luiting2020optimal}. For each parameter (rows), the first column reports the estimate obtained by maximizing the likelihood. The second column contains the standard error, and the third column shows the p-value.}
\begin{tabular}{r||rrr}
  \hline
& Estimate & Std. Error &  p-value \\ 
  \hline
(Intercept) & -0.2754 & 0.4757 & 0.5627 \\ 
  $P^R$ & 0.6082 & 0.3622 & 0.0931 \\ 
  $P^T$ & 0.0918 & 0.3613 & 0.7994 \\ 
  $P^N$ & 0.3738 & 0.5504 &  0.4971 \\ 
  $S$ & 1.0768 & 0.3628 & 0.0030 \\ 
  Age & -0.1381 & 0.1649 &  0.4023 \\ 
  log(PSA) & 1.3855 & 0.2628 & 0.0000 \\ 
  time & 0.0070 & 0.0042 &  0.0915\\  
   \hline
   \end{tabular}
   \label{logitTABLE}
\end{table}

\subsection{A Cross-Validation study}
To evaluate the performance of the proposed algorithm on the real dataset, we additionally perform a leave-one-out cross-validation analysis (LOOCV), according to \cite{gilks1995markov}. We exploit the cross-validation predictive density sets 
$$\{
f(y_i(t)|\log \mathbf{x}^o, \boldsymbol{\pi}^o ,\log \mathbf{y}^o_{it},  \mathbf{z}^{o});\quad  \forall   \ y_i(t) \ \in \ \mathbf{y}^o\},$$
for the PSA levels, where $\log \mathbf{y}^o_{it}$ is the set of observed log-PSA values except for $\log y_i(t)$, and
$$\{f(z_i(t)|\log \mathbf{x}^o, \boldsymbol{\pi}^o ,\log \mathbf{y}^o,  \mathbf{z}^{o}_{it});\quad  \forall   \ z_i(t) \ \in \ \mathbf{z}^o\},$$
for the PET-PSMA examination,  where $\mathbf{z}^o_{it}$ is the set of observed log-PSA values except for $ z_i(t)$.

\begin{figure}[h!]
    \centering
    \includegraphics[scale=0.5]{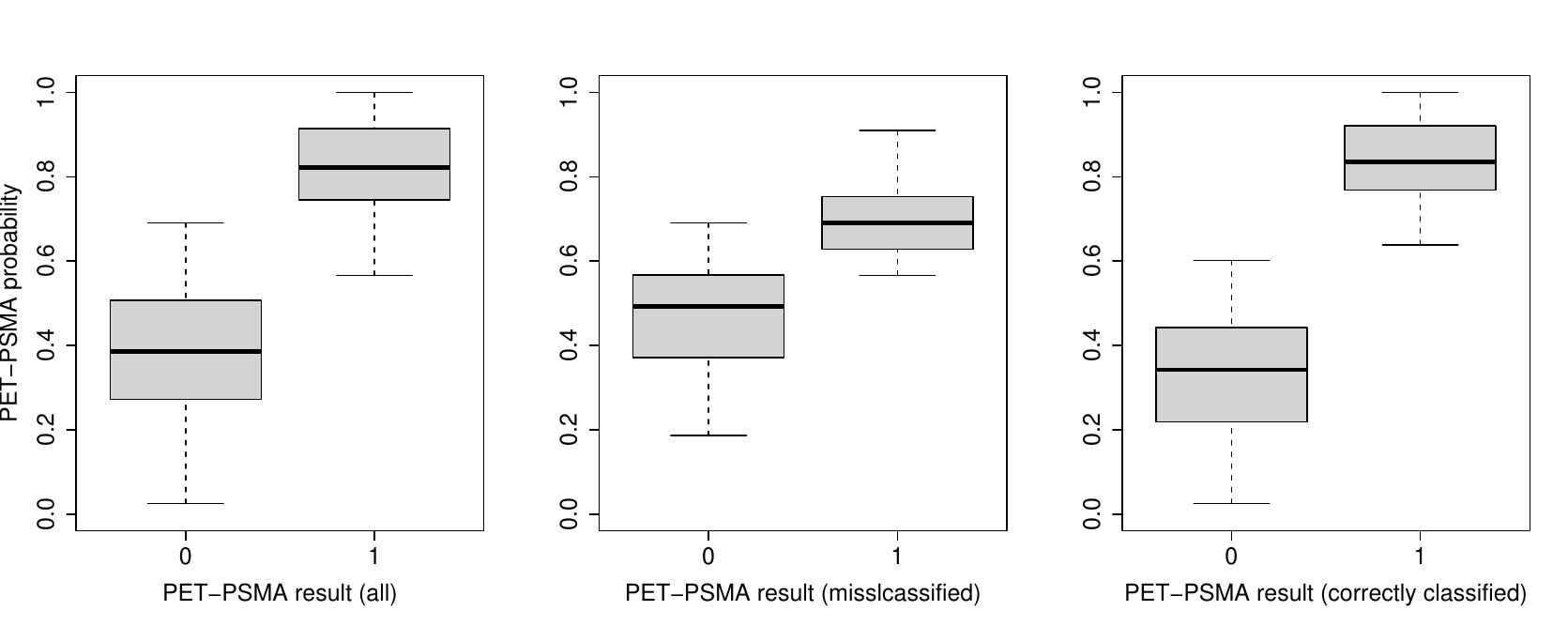}
    \caption{Real Data Application - Predictive probability of positive PET-PSMA results for observed negative (0) and positive (1) examinations of the whole dataset (left panel), of misclassified results only (middle panel), and correctly classified results (right panel), obtained through cross-validation.}
    \label{compare_pet__cv}
\end{figure}

\begin{figure}[h!]
    \centering
    \includegraphics[width=30em]{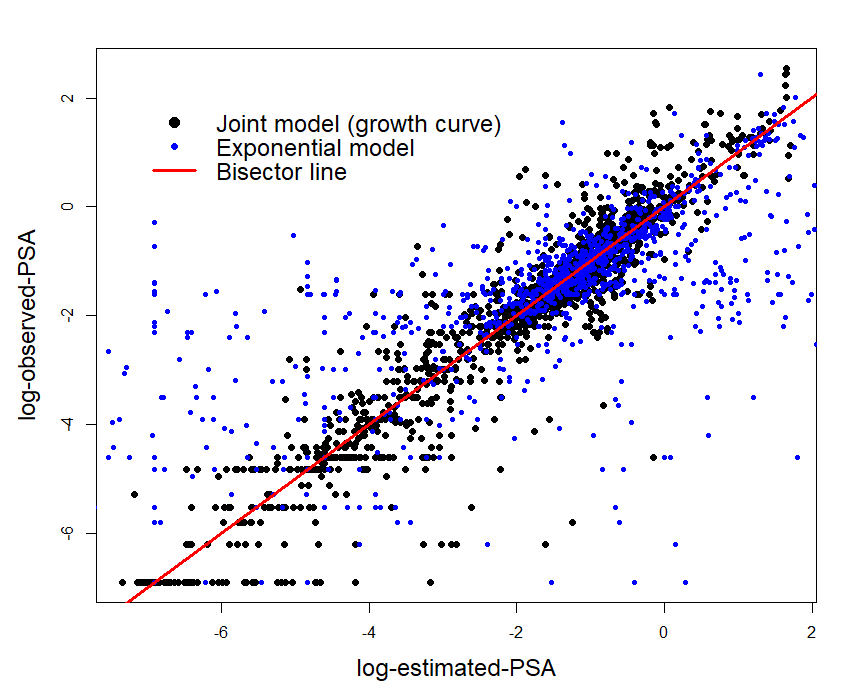}
    \caption{Real Data Application - Posterior mean of the log-psa (on x-axis) compared with the observed log-psa (y-axis): the black points are estimated through our joint model, and the blue points are estimated through the exponential model. The red line is the bisector. Results are obtained through cross-validation.}
    \label{compare_psa__cv}
\end{figure}

For each PSA and PET-PSMA measurement collected, we sample 1000 realizations from the corresponding cross-validation predictive density, estimated through 150.000 iterations of the algorithm (burn-in 100.000, thinning parameter 10). To evaluate the goodness of the prediction we use the accuracy index. The PSA accuracy, i.e. the percentage of predictive $95\%$ CI containing the true values of the corresponding measurement, equals $98.20\%$. On the other side, PET-PSMA accuracy, i.e. the percentage of PET-PSMA results correctly predicted (for each data, the threshold is selected through LOOCV on ROC curves), equals $72.58 \%$.
Moreover, to validate the posterior distribution of $\pi_i(t)$, Figure \ref{compare_pet__cv} shows the probability of positive results across different subsets ($0$ for negative and $1$ for positive) PET-PSMA results. In the left panel, we present the overall dataset, the middle panel focuses on misclassified instances, and, finally, the right panel highlights correctly classified cases.
The plot shows that the posterior median of $\pi_i(t)$, respectively for the outcome $0$ and $1$, is approximately $0.38$ and $0.82$ when considering the entire dataset. However, when we restrict our analysis to the correctly characterized data, these values shift to $ \approx 0.33$ for $0$ and $\approx 0.85$ for $1$, which demonstrates that the right decision can be made with a high degree of confidence.
In the middle panel, when the method encounters challenges in classifying the data, there is a notable increase in uncertainty. The values of $\pi_i(t)$ are close to the threshold $\pi^*$ (potentially around 0.55), confirming that the methods fail when uncertainty is high. This underscores the sensitivity of the model to situations with increased ambiguity and suggests that the uncertainty in predictions increases when the method encounters difficulty in making accurate classifications.
\newline Moreover, we compare the performance and the accuracy of our joint model with results obtained with standard methods, consistently with existing literature practices. 
We conduct leave-one-out cross-validation to assess the performance of the PET-PSMA logistic model (based on  \cite{luiting2020optimal} and presented in Section \ref{sectionROC}), but also the exponential-fit model, a simpler method which predicts the future (and yet unknown) PSA level at time $t_3$ based on pairs of PSA measurements $(PSA_1, PSA_2)$ collected at previous times $t_1$ and $t_2$, utilizing an exponential fit \cite{PSAKinetics,linkpsaexponential}.
Note that the PSA level at the given time $t_3$, is predicted by only using two previous recorded values and not the entire history. While this technique offers predictions, its efficacy is limited compared to our joint model, as shown in Figure \ref{compare_psa__cv}. It illustrates the difference between true and predicted log-PSA levels, estimated with both exponential and joint models. The plot distinguishes predictions made with the exponential model (1647 predictions shown in blue) and our joint model (1671 predictions represented in black), underlying the limited predictive capacity of the exponential basic model, which handles 24 fewer data predictions due to the intrinsic structure of the prediction mechanism (points referring to time $t_1$ and $t_2$ can not be handled). Nevertheless, it is worth noting that the variability of the predicted values with the exponential model is higher than that with the proposed joint model. This highlights the superior predictive performance of the proposed joint model.
\textcolor{black}{Finally, with a standard LOOCV approach, the simple logistic model achieved an accuracy of 65.42$\%$ for predicting PET-PSMA examination outcomes (with a balanced accuracy of 66.27$\%$). In contrast, our joint model improves on these results, reaching an accuracy of 71.25$\%$ and a balanced accuracy of 72.58$\%$. We included balanced accuracy as an additional evaluation metric to address potential errors arising from the imbalance between $0$ and $1$ PET-PSMA results in the dataset, which may be influenced by biological factors (see \cite{ref1} and \cite{ref2}).}

\section{Final remarks and conclusions}\label{sec5}
Correct and quick identification of the locations of possible metastasis in prostate cancer is a challenging open problem. Despite the availability of several new techniques, their calibration is still debated. In particular, the results obtained using the sensitive nuclear examination known as PET-PSMA can be improved if correctly combined with a good estimation of the optimal time to perform the examination. The previous sections contain our proposal on how to estimate the optimal time to perform PET-PSMA which exploits information from the whole history data of each patient. We have introduced the joint model approach, addressing both PSA growth and the probability of a positive PET-PSMA, enriched with random effects to enable predictions for future patients. Our proposal is therefore not just a method for predicting individual PSA growth curve and time to PET-PSMA, but a proposal to drive optimal decisions regarding the patient, which is the core of personalized medicine.
In the paper, after explaining the model structure and the joint approach, we have discussed the optimal time estimation. Finally, the model has been estimated both on simulated and real data. Simulations were used to test the proposed model under challenging settings. 
In particular, we showed that the resurgence changing point time $\tau$ is difficult to estimate, as well as the regressive parameters entering the mean of the random effects. On the other hand, simulations also highlight the adaptability of the method to quite different growth patterns of PSA. The results obtained on real data, for an optimal probability $\pi^*$ selected via LOOCV, with a confidence level of $\rho=95\%$ on the result, give new insights into the model applicability and performance. 
\textcolor{black}{However, the timing of the PET-PSMA examinations not being scheduled a priori is a possible source of bias, as it depends on the natural history of the disease. Although ignoring this component is a common practice, we are currently considering a possible extension of our model that can address this issue if the data and covariates contain enough information.}
On the other hand, both the growth model estimated patterns and the logistic results are easy to interpret and align with clinical evidence. Further research directions could include the implementation of a graphical interface to help clinicians easily exploit the model: any new patient under analysis should be easily included in the database through the interface, enforcing the model accuracy, and enabling good and quick predictions.

\vspace*{1pc}
\noindent {\bf{Author contributions statement}}
M.A. preprocessed the data, developed the model, implemented the code, and wrote and revised the manuscript. 
G.M. developed the model, implemented the code, and wrote the manuscript. 
M.G. formulated the model and revised the manuscript.
S.DL. provided with the data, helped with the clinical interpretation of the results, and revised the manuscript.
\vspace*{1pc}

\noindent {\bf{ORCID}}

\noindent\textit{Martina Amongero} \\
 \url{https://orcid.org/0000-0003-3514-6803}\\
\textit{Gianluca Mastrantonio}\\
\url{https://orcid.org/0000-0002-2963-6729}\\ 
\textit{Mauro Gasparini}\\
 \url{https://orcid.org/0000-0001-8011-4005}\\

\vspace*{1pc}
\noindent {\bf{Acknowledgement}}
\noindent {{The authors thankfully acknowledge HPC@POLITO (\url{http://www.hpc.polito.it}) which provided computational resources. The authors thankfully acknowledge Edoardo Cisero and Angela Pecoraro who partially collected the data.}}

\vspace*{1pc}
\noindent {\bf{Conflict of Interest}}
\noindent {\it{The authors have declared no conflict of interest.}}

\vspace*{1pc}
\noindent {\bf{Data Availability statement}}
The participants of this study did not give written consent for their data to be shared publicly, so due to the sensitive nature of the research supporting data is not available.

\clearpage

\section*{Appendix}\label{appendix-label}
\subsection*{A.1.\enspace Simulations study: additional results}\label{appendix-label-simulation}
\textcolor{black}{In this section, we report additional details on the results obtained in the simulation study of  Section \ref{sec:sim}, with a particular focus on the global parameters.
  We remind the reader that Section \ref{sec:sim} is divided into two parts, the first one analyzing a single dataset (S1),  while the second one is about a full simulation study composed of 4 scenarios, with 100 datasets each (S2).
  In Table \ref{global-1dataset}, we report the global parameter estimates for the single scenario presented in Subsection \ref{onedata}.
In Table \ref{scenarios_table}, we report all the parameter values used to simulate each of the datasets referring to four scenarios (see subsection \ref{sec:simfull}). Finally, in Tables \ref{tab:simulation_full2_new} and \ref{tab:simulation_full2a}, we report the global parameters estimates and some related diagnostics (four simulated scenarios, with 100 repetitions each presented in \ref{sec:simfull}).}

\begin{table}[H]
\centering
\caption{S1 - Global parameters describing the PSA level. The table shows the limits of the 95\% CIs (2.5\% and 97.5\%), the posterior mean (Mean), the true value used to simulate the data (Real), and the value of the $\hat{R}$ statistic.}\label{table_simuation_glob}
\begin{tabular}{c||ccccc}
\hline
 Parameter &2.5\%  & Mean & 97.5\% & Real  & $\hat R$  \\ \hline   

   $\alpha_{\mu}[1]$ & 0.91 & 0.99 & 1.06 & 1.00 & 1.00 \\ 
   $\alpha_{\mu}[2]$ & 0.01 & 0.07 & 0.13 & 0.10 & 1.00 \\ 
   $\alpha_{\mu}[3]$ & 0.24 & 0.29 & 0.35 & 0.30 & 1.00 \\ 
   $\alpha_{\mu}[4]$ & 0.45 & 0.51 & 0.57 & 0.50 & 1.00\\ 
   $\alpha_{\mu}[5]$ & 0.13 & 0.19 & 0.25 & 0.20 & 1.00 \\ 
   $\alpha_{\mu}[6]$ & 0.09 & 0.15 & 0.21 & 0.10 & 1.00 \\ 
   $\alpha_{\gamma}[1]$ & -1.06 & -0.95 & -0.84 & -1.00 & 1.00 \\ 
   $\alpha_{\gamma}[2]$ & -0.11 & -0.04 & 0.03 & 0.00 & 1.00 \\ 
   $\alpha_{\gamma}[3]$ & -0.04 & 0.02 & 0.09 & 0.00 & 1.00 \\ 
   $\alpha_{\gamma}[4]$ & -0.04 & 0.04 & 0.11 & 0.00 & 1.00 \\ 
   $\alpha_{\gamma}[5]$ & -0.08 & -0.01 & 0.06 & 0.00 & 1.00 \\ 
   $\alpha_{\gamma}[6]$ & -0.14 & -0.07 & 0.00 & 0.00 & 1.00 \\    
   $\beta_1$ & 2.74 & 4.94 & 8.54 & 4.00 & 1.00 \\
   $\beta_2$ & 0.50 & 0.84 & 1.32 & 0.50 & 1.00 \\ 
   $\alpha_{\beta}[1]$ & -6.45 & 6.77 & 21.14 & 1.00 & 1.00 \\
   $\alpha_{\beta}[2]$ & -0.38 & 2.51 & 5.81 & 1.00 & 1.00 \\ 
   $\alpha_{\beta}[3]$ & -1.82 & 1.03 & 4.06 & 1.00 & 1.00 \\ 
   $\alpha_{\beta}[4]$ & -2.03 & 1.05 & 4.33 & 0.50 & 1.00 \\ 
   $\alpha_{\beta}[5]$ & -4.71 & -1.18 & 1.43 & -0.50 & 1.00 \\ 
   $\alpha_{\beta}[6]$ & -1.24 & -0.79 & -0.46 & -0.50 & 1.00 \\ 
   $\omega_{\mu}$ & 0.09 & 0.10 & 0.13 & 0.100 & 1.00 \\ 
   $\omega_{\gamma}$  & 0.06 & 0.10 & 0.14 & 0.10 & 1.00 \\    
   $\psi_a$ & 4.81 & 5.64 & 6.51 & 5.70 & 1.00 \\ 
   $\omega_a$ & 0.85 & 1.21 & 1.73 & 1.00 & 1.00 \\ 
   $a_{\sigma^2}$ & 2.00 & 2.24 & 3.69 & 3.00 & 1.00 \\ 
   $b_{\sigma^2}$ & 3.69 & 5.28 & 9.49 & 5.00 & 1.00 \\
   \hline
   \end{tabular}
\label{global-1dataset}

\end{table}

\begin{table}[H]
    \centering
    \caption{S2 - Parameters used to simulate the datasets for each scenario}
     \setlength{\tabcolsep}{3pt} 
    \begin{tabular}{c||cccccc|cccccc}
    \hline
                   &     &     &  $\alpha_{\mu}$   &     &     &     &   &  & $\alpha_{\gamma}$&  & \\ \hline
       \small{Scenario 1} & 0.5 & 0.1 & 0.3 & 0.5 & 0.2 & 0.1 &  -0.5 & -0.1 & -0.1 & -0.1 & -0.1 & -0.1   \\ 
        \small{Scenario 2}& 0.5 & 0.1 & 0.3 & 0.5 & 0.2 & 0.1 &  -0.5 & -0.1 & -0.1 & -0.1 & -0.1 & -0.1   \\ 
         \small{Scenario 3} & 0.1 & 0.3 & 0.3 & 0.5 & 0.2 & 0.5 &  -0.1 & -0.5 & -0.5 & -0.5 & -0.5 & -0.5 \\ 
          \small{Scenario 4}& 0.5 & 0.1 & 0.3 & 0.5 & 0.2 & 0.1 &  -0.5 & -0.1 & -0.1 & -0.1 & -0.1 & -0.1   \\  \hline
                   &     &     &  $\alpha_{\beta}$   &     &     &   &$\beta_1$  & $\beta_2$   & ($\psi_a$; $\omega_a$) & $\omega_{\mu,\gamma}$&  $a_{\sigma^2}$ & $b_{\sigma^2}$\\ \hline 
       \small{Scenario 1}& 0.5 & 1.0 & 1.0 & 0.5 & -0.5 & -0.5 & 1.0 & 2.0  & (2.48 ; 1) & 0.1 & 3 & 5 \\ 
         \small{Scenario 2}& 0.5 & 1.0 & 1.0 & 0.5 & -0.5 & -0.5 &1.0 & 2.0  & (2.48 ; 1)  & 0.1 & 3 & 15 \\ 
         \small{Scenario 3} & 0.5 & 0.5 & 0.5 & 0.1 & -0.1 & -0.1 & 0.2 & 4.0 & (2.48 ; 1)  & 0.5 & 2 & 7 \\ 
        \small{Scenario 4}& 0.5 & 1.0 & 1.0 & 0.5 & -0.5 & -0.5 & 1.0 & 2.0  & (2.48 ; 1)  & 0.1 & 3 & 5 \\  \hline
    \end{tabular}
    \label{scenarios_table}
\end{table}

\begin{table}[H]
    \centering
    \caption{S2 -   First set of global parameters. For each scenario (columns) and parameter (rows), the first row reports the percentage of correctly estimated parameters. The second row shows the empirical quantiles at levels \([2.5 \%, 97.5 \%]\) and the sample mean (the value in the center) of the distribution of the confidence interval (CI) widths. The third row displays the empirical quantiles at levels \([2.5 \%, 97.5 \%]\) and the sample mean (the value in the center) of \(\hat{R}\). All components of the table are computed across 100 datasets.}\label{tab:simulation_full1}
    \setlength{\tabcolsep}{2pt} 
    \begin{tabular}{r|c||cccc}
    \hline
      {Parameter}&  & {Scenario 1 } & {Scenario 2} & {Scenario 3} & {Scenario 4} \\ \hline 
        &\%  &  92$\%$ &  83$\%$ & 90$\%$ & 92$\%$ \\
$\alpha_{\mu}[1]$ & $ CI_{W} $  &  \small{[0.012 0.16 0.19]}&  \small{[0.15 0.21 0.29]}& \small{[0.52  0.64  0.79]} &  \small{[0.12 0.15 0.19]}\\
&$  CI_{\hat R} $   &  \small{[1.00 1.01 1.05]}&  \small{[1.00 1.03 1.15]}& \small{[1.00 1.00 1.02]} &  \small{[1.00  1.01  1.05]}\\
\hline

&  \% &  97$\%$& 94$\%$ & 92$\%$ & 96$\%$ \\
$\alpha_{\mu}[2]$ &$ CI_{W}$    &  \small{[0.09 0.12 0.15]} & \small{[0.06 0.14 0.18]}& \small{[0.43  0.50  0.59]} & \small{[0.09  0.12  0.15]}\\
&$ CI_{\hat R} $   &  \small{[1.00 1.01 1.04]}&  \small{[1.00 1.04 1.18]}&  \small{[1.00 1.01 1.03]} &  \small{[1.00 1.01 1.04]}\\
\hline

&  \%  &  98$\%$ & 92$\%$ & 97$\%$ & 98$\%$\\
$\alpha_{\mu}[3]$&$ CI_{W}$    & \small{[0.10  0.12 0.14]} &  \small{[0.08 0.14 0.17]} &  \small{[0.42  0.50  0.56]} & \small{[0.10  0.12  0.14]}\\
&$CI_{\hat R}  $   &  \small{[1.00 1.01 1.06]}&  \small{[1.00 1.03 1.16]}&  \small{[1.00 1.01 1.04]}&  \small{[1.00 1.01 1.07]}\\

\hline

& \% &  97$\%$& 95$\%$& 92$\%$ &  97$\%$ \\
$\alpha_{\mu}[4]$  &$ CI_{W}$    &  \small{[0.10 0.12 0.15]}& \small{[0.08 0.14 0.18]}&  \small{[0.43  0.50  0.58]} & \small{[0.10  0.12  0.15]}\\
&$ CI_{\hat R} $   &   \small{[1.00 1.01 1.05]}&  \small{[1.00 1.03 1.14]}&  \small{[1.00 1.01 1.02]} &  \small{[1.00 1.01 1.05]}\\

\hline
& \% &  94$\%$ & 88$\%$ & 96$\%$ & 93$\%$ \\
$\alpha_{\mu}[5]$ &$ CI_{W}$   &  \small{[0.009 0.12 0.14]}& \small{[0.06 0.14 0.18]}&  \small{[0.44  0.50  0.59]} & \small{[0.09  0.12  0.14]}\ \\
&$CI_{\hat R}  $   &  \small{[1.00 1.01 1.06]}&  \small{[1.00 1.03 1.20]}& \small{[1.00 1.01 1.03]}&  \small{[1.00 1.01 1.06]}\\

\hline

&  \% &  94$\%$& 95$\%$& 94$\%$ & 94$\%$\\
$\alpha_{\mu}[6]$ &$ CI_{W}$    & \small{[0.09 0.12 0.15]}& \small{[0.06 0.14 0.18]}& \small{[0.43  0.50  0.57]} & \small{[0.09  0.12  0.15]}\\
&$CI_{\hat R}  $   &   \small{[1.00 1.01 1.05]}&  \small{[1.00 1.03 1.18]}&  \small{[1.00 1.01 1.03]} &  \small{[1.00 1.01 1.05]}\\

\hline

& \%&  96$\%$ & 85$\%$ & 98$\%$ & 96$\%$ \\
$\alpha_{\gamma}[1]$&$CI_{W} $  &  \small{[0.28 0.37 0.50]}& \small{[0.01 0.41 0.72]}&  \small{[0.61 0.77  0.94]} & \small{[0.28  0.37  0.5]}\\
&$CI_{\hat R}  $   &  \small{[1.00 1.07 1.36]}&  \small{[1.00 1.17 1.92]}&  \small{[1.00 1.01 1.04]} &  \small{[1.00 1.07 1.36]}\\

\hline
&\% &  99$\%$ & 87$\%$ & 97$\%$ &  99$\%$\\
$\alpha_{\gamma}[2]$&$ CI_{W}$ &  \small{[0.18 0.23 0.28]} & \small{[0.00 0.24 0.35]}& \small{[0.44  0.55  0.64]} & \small{[0.18  0.23  0.28]}\\
&$CI_{\hat R}  $   &   \small{[1.00 1.01 1.08]}&  \small{[1.00 1.08 1.62]}&  \small{[1.00 1.00 1.02]} &  \small{[1.00 1.01 1.08]}\\

\hline

&\% &  97$\%$ &86$\%$ & 94$\%$ &  97$\%$ \\
$\alpha_{\gamma}[3]$&$CI_{W} $ &  \small{[0.18,  0.23 , 0.29]} &\small{[0.00 0.25 0.36]}& \small{[0.43  0.54  0.63]} & \small{[0.18  0.23  0.29]}\\
&$ CI_{\hat R} $   &   \small{[1.00 1.01 1.06]}&  \small{[1.00 1.07 1.68]}&  \small{[1.00 1.00 1.03]} &  \small{[1.00 1.01 1.06]}\\

\hline
& \%&  98$\%$& 86$\%$& 96$\%$ &  98$\%$ \\
$\alpha_{\gamma}[4] $&$CI_{W} $ &  \small{[0.18 0.24 0.31]}& \small{[0.00 0.26 0.43]}& \small{[0.44  0.55  0.63]} & \small{[0.18  0.23  0.31]}\\\
&$CI_{\hat R}  $   &  \small{[1.00 1.02 1.11]}&  \small{[1.00 1.08 1.55]}&  \small{[1.00 1.00 1.02]} &  \small{[1.00 1.02 1.11]}\\

\hline
&\%&  99$\%$& 87$\%$& 95$\%$ &  99$\%$ \\
$\alpha_{\gamma}[5] $ &$ CI_{W}$ &  \small{[0.18 0.23 0.28]}& \small{[0.00 0.25 0.37]}&  \small{[0.43  0.54  0.62]} & \small{[0.18  0.23  0.28]}\\
&$CI_{\hat R}  $   &   \small{[1.00 1.01 1.06]}&  \small{[1.00 1.10 1.79]}&  \small{[1.00 1.00 1.02]}&  \small{[1.00 1.01 1.06]}\\

\hline
& \%&  99$\%$ & 88$\%$ & 94$\%$ &  99$\%$ \\
$\alpha_{\gamma}[6]$&$CI_{W} $ &   \small{[0.18 0.23 0.29]}& \small{[0.00 0.24 0.39]}&  \small{[0.44  0.55  0.63]} & \small{[0.18  0.23  0.29]}\\
&$CI_{\hat R}  $   &   \small{[1.00 1.01 1.04]}&  \small{[1.00 1.07 1.65]}&  \small{[1.00 1.00 1.02]} &  \small{[1.00 1.01 1.04]}\\

     \hline

    \end{tabular}
     \label{tab:simulation_full2_new}
\end{table}

\begin{table}[H]
    \centering
    \caption{S2 -  Second set of global parameters. For each scenario (columns) and parameter (rows), the first row reports the percentage of correctly estimated parameters. The second row shows the empirical quantiles at levels \([2.5 \%, 97.5 \%]\) and the sample mean (the value in the center) of the distribution of the confidence interval (CI) widths. The third row displays the empirical quantiles at levels \([2.5 \%, 97.5 \%]\) and the sample mean (the value in the center) of \(\hat{R}\). All components of the table are computed across the 100 datasets.}\label{tab:simulation_full1_2}
    \setlength{\tabcolsep}{2pt} 
    \begin{tabular}{r|c||cccc}
    \hline
      \

&       \%   &  43$\%$ & 49$\%$ & 5$\%$ & 43$\%$\\
$\beta_1$ &$ CI_{W}$           &   \small{[4.68	13.43 22.82]}&  \small{[4.39 12.94 22.85]}& \small{[8.24 21.46 29.51]} & \small{[4.68 13.50 23.86]}\\\
&$CI_{\hat R}  $   &  \small{[1.00 1.04 1.13]}&  \small{[1.00 1.04 1.21]}& \small{[1.00 1.11 1.51]} &  \small{[1.00 1.04 1.13]}\\
\hline

& \%&  56$\%$ & 53$\%$ & 84 $\%$ & 56$\%$  \\
$\beta_2$&$ CI_{W} $  &  \small{[0.54 2.24 9.95]}&  \small{[0.56 2.29 9.18]}&  \small{[3.18  7.33 11.62]} &\small{[0.54  2.30  9.95]}\\
&$ CI_{\hat R} $   &  \small{[1.00 1.02 1.12]}&  \small{[1.00 1.02 1.10]}& \small{[1.00 1.02 1.10]} &  \small{[1.00 1.02 1.12]}\\

\hline

& \% &  100$\%$ & 100$\%$& 100$\%$ & 100$\%$\\
$\alpha_{\beta}[1]$&$ CI_{W}$  &  \small{[24.76	35.22 39.41]}& \small{[25.94 35.14 39.54]}&  \small{[37.56 38.71 39.87]} & \small{[24.76 35.23 39.41]}\\
&$ CI_{\hat R} $   &  \small{[1.00 1.00 1.03]}&  \small{[1.00 1.00 1.03]}& \small{[1.00 1.00 1.00]} &  \small{[1.00 1.00 1.03]}\\
\hline

&  \% &  95$\%$ & 94$\%$ & 100$\%$ &  95$\%$ \\
$\alpha_\beta[2]$ &$ CI_{W}$    &   \small{[4.43 15.14 30.35]}& \small{[4.51 14.92 31.20]}& \small{[22.46 31.55 36.33]} & \small{[4.43 15.28 30.61]}\\
&$CI_{\hat R}  $   &  \small{[1.00 1.01 1.03]}&  \small{[1.00 1.01 1.03]}& \small{[1.00 1.00 1.01]} &  \small{[1.00 1.01 1.03]}\\
\hline

&    \%  &  92$\%$ & 92$\%$& 100$\%$ &  92$\%$\\
$\alpha_{\beta}[3]$&$CI_{W} $  &  \small{[44.21	15.29	29.43]}& \small{[4.28 15.02 30.36]}&  \small{[21.05 31.68 35.40]} & \small{[4.21 15.41 30.32]}\\
&$ CI_{\hat R} $   &  \small{[1.00 1.01 1.04]}&  \small{[1.00 1.01 1.04]}& \small{[1.00 1.00 1.02]} &  \small{[1.00 1.01 1.04]}\\

\hline

& \% &  94$\%$ &  92$\%$& 100$\%$ &  94$\%$ \\
$\alpha_{\beta}[4]$&$ CI_{W}$  &   \small{[4.37	14.97 30.30]}&  \small{[4.22 14.71 30.51]}&  \small{[22.57 31.75 35.79]} & \small{[4.37 15.09 30.30]}\\
&$ CI_{\hat R} $   & \small{[1.00 1.01 1.04]}&  \small{[1.00 1.01 1.04]}& \small{[1.00 1.00 1.01]} &  \small{[1.00 1.01 1.04]}\\

\hline

& \%&  98$\%$ & 97$\%$ & 100$\%$ &  98$\%$\\
$\alpha_{\beta}[5]$ &$ CI_{W} $  &  \small{[4.19 14.79 30.36]}&  \small{[4.09 14.43 30.64]}&  \small{[22.07 31.54 35.83]} & \small{[4.19 14.91 30.92]}\\
&$ CI_{\hat R} $   & \small{[1.00 1.01 1.03]}&  \small{[1.00 1.01 1.03]}& \small{[1.00 1.00 1.01]} &  \small{[1.00 1.01 1.03]}\\

\hline

& \% &  36$\%$ & 45$\%$ &  94$\%$ &  36$\%$ \\
$\alpha_{\beta}[6]$&$CI_{W} $  &  \small{[0.58	2.14 6.90]}& \small{[0.61 2.05 6.19]}& \small{[1.35  2.94  4.59]} & \small{[0.58  2.18  6.90]}\\
&$CI_{\hat R}  $   &  \small{[1.00 1.03 1.15]}&  \small{[1.00 1.02 1.14]}& \small{[1.00 1.01 1.08]} &  \small{[1.00 1.03 1.15]}\\

\hline
&   \%  &  91$\%$  & 84$\%$ & 99$\%$ &  91$\%$  \\
$\omega_{\mu} $ &$ CI_{W} $      &   \small{[0.04 0.05 0.06]}&  \small{[0.05 0.07 0.1]}&  \small{[0.17  0.20  0.23]} &\small{[0.04  0.05  0.06]}\\
&$CI_{\hat R}  $   &  \small{[1.00 1.02 1.10]}&  \small{[1.00 1.07 1.29]}& \small{[1.00 1.01 1.05]} &  \small{[1.00 1.02 1.10]}\\
\hline

&   \%&  62$\%$ & 77$\%$ & 96$\%$ &  63$\%$\\
$\omega_{\gamma}$ &$CI_{W} $    &  \small{[0.10 0.14 0.25]}&  \small{[0.00 0.18 0.31]}&  \small{[0.17  0.22  0.26]} &\small{[0.10  0.14  0.25]}\\
&$CI_{\hat R}  $   & \small{[1.00 1.03 1.19]}&  \small{[1.00 1.10 1.54]}& \small{[1.00 1.00 1.02]} &  \small{[1.00 1.03 1.19]}\\

\hline
&   \%          &  93$\%$ & 88$\%$ & 99$\%$ &  93$\%$ \\
$\psi_a$&$ CI_{W}$             &   \small{[1.11 1.72 3.04]}&  \small{[1.33 2.12 3.47]}& \small{[0.97  1.35  1.96]} &\small{[1.11  1.72  3.04]}\\
&$CI_{\hat R}  $   &  \small{[1.00 1.15 1.65]}&  \small{[1.00 1.28 2.29]}& \small{[1.00 1.06 1.30]} &  \small{[1.00 1.15 1.64]}\\

\hline
&   \%      &  96$\%$& 96$\%$ & 95$\%$ &  96$\%$\\
$\omega_a $ &$ CI_{W}$          &  \small{[0.68 1.04 1.72]}& \small{[0.69 1.30 2.68]}& \small{[0.77 1.20 1.77]} &\small{[0.68  1.05  1.72]}\\
&$CI_{\hat R}  $   &  \small{[1.00 1.07 1.38]}&  \small{[1.00 1.10 1.42]}& \small{[1.00 1.07 1.41]} &  \small{[1.00 1.07 1.38]}\\
\hline

&    \%  &  96$\%$ & 96$\%$ & 100$\%$ &  96$\%$\\
$a_{\sigma^2}$ &$CI_{W} $       &   \small{[0.96 2.92 6.54]}& \small{[1.08 3.43 8.48]}& \small{[0.21  0.81  2.28]} &\small{[0.96  2.90  6.54]}\\
&$CI_{\hat R}  $   &  \small{[1.00 1.01 1.04]}&  \small{[1.00 1.02 1.10]}& \small{[1.00 1.00 1.01]} &  \small{[1.00 1.01 1.03]}\\
\hline

&   \%  &  99$\%$& 92$\%$& 81$\%$ &  99$\%$\\
$b_{\sigma^2} $ &$CI_{W} $      &  \small{[3.44 8.80 18.76]}& \small{[8.94 28.92 71.10]}& \small{[3.21  5.64 13.47]} &\small{[3.44  8.75 18.76]}\\
&$CI_{\hat R}  $   & \small{[1.00 1.01 1.08]}&  \small{[1.00 1.02 1.19]}& \small{[1.00 1.00 1.02]} &  \small{[1.00 1.01 1.07]}\\
     \hline

    \end{tabular}
     \label{tab:simulation_full2a}
\end{table}

\clearpage

\subsection*{A.2 \enspace Additional simulation studies (S3)}\label{appendix-label-simulation-realistic}
\textcolor{black}{In this section, we provide additional insights into the model applicability and performance by analyzing 100 simulated datasets. These datasets were generated using parameters derived from the real data, specifically the mean values of the posterior distribution as reported in Tables  \ref{table_mu}-\ref{table_gamma}-\ref{table_logit}-\ref{table_global}. The aim is to assess model performance in a realistic scenario, with data simulated in accordance with the joint model under analysis, serving as a sensitivity check of the real data analysis. Consistent with the structure of the main text, individual parameter results are presented in Table 13, and global parameter results in Table 14. Overall, results are highly satisfactory, with high percentages, narrow confidence intervals, and mean $\hat R$  indices mostly below $1.1$.}

\begin{table}[t]
\centering
\caption{S3 -  Individual parameters. The first column reports the percentage of correctly estimated parameters. The second column shows the empirical quantiles at levels \([2.5 \%, 97.5 \%]\) and the sample mean (the value in the center) of the distribution of the confidence interval (CI) widths. The third row displays the empirical quantiles at levels \([2.5 \%, 97.5 \%]\) and the sample mean (the value in the center) of \(\hat{R}\). All components of the table are computed across the 80 individuals and 100 datasets.}
\label{table_simuation_realistic_ind}  
\begin{tabular}{c||ccc}
\hline
 Parameter & Percentage   & $CI_{W}$ & $ CI_{\hat R}$   \\ 
 \hline
 $\log(\lambda_i)$  & 93\% & [0.38  1.64  5.34] & [1.00  1.02  1.15] \\
 $\tau_i$           & 82\%  & [0.82 29.07 80.20] & [1.00  1.17  2.21]\\
  $\log(\gamma_i)$  & 91\%  & [0.00 1.01 5.18] & [1.00  1.01  1.07]\\
 $\log(\mu_i)$      & 86\%    & [0.00  0.05 0.10] & [1.00  1.03  1.20] \\
 $\log(a_i)$        & 93\%  & [0.32  1.06  2.61] & [1.00  1.02  1.15] \\
 $\sigma^2_i$       & 91\%   & [0.17  1.37  1.38] & [1.00  1.01  1.09]\\
  \hline
\end{tabular}
\end{table}

 \begin{table}[t]
\centering
\caption{S3 - Global parameters describing the PSA level and PET-PSMA results. 
The first column reports the percentage of correctly estimated parameters. The second column shows the empirical quantiles at levels \([2.5 \%, 97.5 \%]\) and the sample mean (the value in the center) of the distribution of the confidence interval (CI) widths. The third columns displays the empirical quantiles at levels \([2.5 \%, 97.5 \%]\) and the sample mean (the value in the center) of \(\hat{R}\). All components of the table are computed across the 100 scenarios.} \label{table_simuation_realistic_glob}
\begin{tabular}{c||ccc}
  \hline
 Parameter &  Percentage & CI$_{W}$  & CI$_{\hat R}$\\ 
  \hline
 $\alpha_{\mu}[1]$ &  89$\%$ & [2.41  3.01  3.72]    &   [1.00  1.06  1.34] \\ 
 $\alpha_{\mu}[2]$ &  92$\%$ & [1.96 2.37  2.87]     &   [1.00  1.04 1.19] \\ 
  $\alpha_{\mu}[3]$ &  88$\%$ & [2.03  2.43  2.89]   &    [1.00  1.05  1.20] \\ 
 $\alpha_{\mu}[4]$ &  94$\%$ & [1.85  2.38  2.98]    &    [1.00  1.05  1.23]\\ 
  $\alpha_{\mu}[5]$ &  93$\%$ & [1.99  2.41  2.83]   &    [1.00  1.04 1.21] \\ 
 $\alpha_{\mu}[6]$ &  95$\%$ & [1.93  2.41  2.92]    &    [1.00  1.04  1.19]\\ 
$\alpha_{\gamma}[1]$ &  88$\%$ & [0.85  1.18  1.54]  &    [1.00  1.06  1.28] \\ 
$\alpha_{\gamma}[2]$ &  92$\%$ & [0.57  0.75  0.96]  &    [1.00  1.05  1.22] \\
$\alpha_{\gamma}[3]$&  90$\%$ &  [0.52  0.76  0.98]  &    [1.00  1.07  1.30] \\ 
$\alpha_{\gamma}[4]$ &  93$\%$ & [0.57  0.77  0.99] &     [1.00  1.08  1.35] \\ 
$\alpha_{\gamma}[5]$ &  92$\%$ & [0.58  0.77  0.95]  &    [1.00  1.06  1.36] \\ 
$\alpha_{\gamma}[6]$&  91$\%$ & [0.28  0.38  0.48]   &    [1.00  1.06  1.29]\\ 
$\alpha_{\gamma}[7]$ &  88$\%$ & [0.58 0.77  0.99]  &     [1.00  1.05  1.24] \\ 
$\alpha_{\gamma}[8]$ &  97$\%$& [0.57  0.79  0.99]   &    [1.00  1.08  1.32]\\ 
$\alpha_{\gamma}[9]$ &  89$\%$& [0.57  0.79  1.00]  &    [1.00  1.07  1.38] \\
 $\alpha_{\gamma}[10]$ &  95$\%$ & [0.54  0.75  0.94] &    [1.00  1.05 1.27] \\ 
 $\alpha_{\gamma}[11]$ &  87$\%$ & [0.56 0.77  0.97] &    [1.00  1.07  1.28] \\ 
 $\beta_1$ &  94$\%$ & [1.18  1.69  3.03]            &    [1.00 1.03  1.23]\\ 
 $\beta_2$  &   93$\%$& [0.01  0.02  0.02]           &   [1.00 1.01  1.06] \\ 
 $\alpha_{\beta}[1]$ &  90$\%$ & [3.31  4.14  5.81]  &    [1.00 1.01  1.08]\\
$\alpha_{\beta}[2]$ &  91$\%$ & [1.50  1.76  2.22]   &    [1.00  1.01  1.03] \\ 
$\alpha_{\beta}[3]$ &  96$\%$ & [1.48  1.74  2.17]   &    [1.00  1.01  1.04] \\ 
$\alpha_{\beta}[4]$ &  95$\%$ & [1.49  1.80 2.30]    &    [1.00  1.01  1.04] \\ 
$\alpha_{\beta}[5]$&  94$\%$ & [1.53  1.83  2.26]    &    [1.00  1.01  1.03]\\ 
$\alpha_{\beta}[6]$&  92$\%$ & [0.75  0.88 1.11]     &    [1.00  1.01 1.03] \\ 
 $\omega_{\mu}$ &  86$\%$ & [0.89 1.05  1.34]        &    [1.00 1.11 1.56]  \\ 
 $\omega_{\gamma}$ &   63$\%$ & [0.25  0.37  0.48]    &   [1.00 1.12 1.63] \\ 
  
   $\psi_a$ &   99$\%$ & [0.20  0.22  0.28]           &   [1.00  1.01  1.04] \\ 
   $\omega_a$ &  93$\%$ & [0.15  0.17  0.21]         &    [1.00 1.02 1.12] \\ 
   $a_{\sigma^2}$ &  100$\%$ & [0.01  0.04  0.11]     &   [1.00  1.00  1.00] \\ 
   $b_{\sigma^2}$ &   78$\%$ & [0.05  0.06  0.08]     &   [1.00  1.01  1.05] \\ 
   \hline
\end{tabular}
\end{table}

\clearpage

\subsection*{A.3.\enspace Real Data Application: additional results}\label{appendix-label-real}
In this part of the Appendix, we report additional details of the real data analysis, presented in Section \ref{sec:real}.
We first present, in Table \ref{summary-cov}, a summary of the real dataset. Then, we report the details of the three models we fit, explaining the different combinations of covariates we use, in Table \ref{table_simuation_global}.

\begin{table}[t]
    \centering
    \caption{Real Data Application - A descriptive table of the covariates is presented. The covariates are divided into dichotomous and continuous variables, with summary statistics reported for each.} \label{tab:my_label_sum}
    \begin{tabular}{c||c|c|cc}
        \hline
     Binary Covariate & Number of 0& Number of 1 & Missing values\\ 
     \hline
      $P^R$  & 120  & 67 & - \\
      $P^T$  & 67 & 120 & -  \\
      $P^N$   & 112 & 23 & 52  \\
      S   &  66 & 121 &  - \\
      OA   & 163 & 24 & -  \\
      OS   &  156 & 31 &-  \\
      RA   & 163 &  24 &  - \\
     RS   &  155 & 32 & -  \\
     L    &  51& 136& -   \\
     \hline
      Continuous covariate & Mean-SD  & $2.5\%-97.5\%$ Quantile  & Missing values\\
          \hline
      Age & 71.32 - 7.81  & [53.00, 84.43]  &  -\\ 
     
         \hline
    \end{tabular} \label{summary-cov}
\end{table}

\begin{table}[t]
\center
\caption{Real Data Application - The table shows the covariates used in the three different parametrization of the proposed model used to analyze the data. A {$\Box$} indicates that the variable is used to model the mean of  $\mu$. A {$\triangle$} indicates that the variable is used to model the mean of $\gamma$. A {$\times$} indicate  that the variable is used in the logistic regression.  For each setting the associated  WAIC is reported.}

\label{table_simuation_global}
\begin{center}
\setlength{\tabcolsep}{0.3mm}
{\begin{tabular}{c||cccccccccc||c}
\hline
   Model & OA & RA & OS & RS & L & $P^R$ & $P^T$ & $P^N$ & S & A &    WAIC \\ \hline
    1
  & {{$\Box$}}{{$\triangle$}}{{$\times$}}  
  & {{$\Box$}}{{$\triangle$}}{{$\times$}} 
  & {{$\Box$}}{{$\triangle$}}{{$\times$}} 
  & {{$\Box$}}{{$\triangle$}}{{$\times$}} 
  & {{$\Box$}}{{$\triangle$}}{{$\times$}} 
  & {{$\Box$}}{{$\triangle$}}{{$\times$}} 
  & {{$\Box$}}{{$\triangle$}}{{$\times$}} 
  & {{$\Box$}}{{$\triangle$}}{{$\times$}} 
  & {{$\Box$}}{{$\triangle$}}{{$\times$}} 
  & {{$\Box$}}{{$\triangle$}}{{$\times$}}  
  & 1230819 \\
 
   2
  & {{$\Box$}}{{$\triangle$}}   
  & {{$\Box$}}{{$\triangle$}}  
  & {{$\Box$}}{{$\triangle$}}  
  & {{$\Box$}}{{$\triangle$}}  
  & {{$\Box$}}{{$\triangle$}}  
  & {{$\triangle$}}{{$\times$}} 
  & {{$\triangle$}}{{$\times$}} 
  & {{$\triangle$}}{{$\times$}} 
  & {{$\triangle$}}{{$\times$}} 
  & {{$\triangle$}}{{$\times$}} 
 & 1093648 \\

    3
  & {{$\Box$}} 
  & {{$\Box$}} 
  & {{$\triangle$}} 
  & {{$\triangle$}} 
  & {{$\Box$}}{{$\triangle$}} 
   & {{$\triangle$}}{{$\times$}} 
  & {{$\triangle$}}{{$\times$}} 
  & {{$\triangle$}}{{$\times$}} 
  & {{$\triangle$}}{{$\times$}} 
  & {{$\triangle$}}{{$\times$}} 
  & 1205765 \\
\hline
\end{tabular}
}
\end{center}
\end{table}

\newpage

\vspace{2em}

\vspace{-2em}
\vspace{-2em}

\vspace{-2em}

\vspace{-2em}

\vspace{-2em}

\end{document}